\documentclass[sigconf,nonacm]{acmart}

\usepackage{subfigure}
\usepackage{siunitx}
\usepackage{enumitem}
\usepackage{xurl}

\usepackage{algorithm}
\usepackage[noend]{algpseudocode}
\let\ReturnInline\Return
\renewcommand{\Return}{\State\ReturnInline}
\algrenewcommand\algorithmicrequire{$\rhd$}
\algrenewcommand\algorithmicensure{$\square$}

\AtBeginDocument{%
  \providecommand\BibTeX{{%
    \normalfont B\kern-0.5em{\scshape i\kern-0.25em b}\kern-0.8em\TeX}}}

\setcopyright{acmcopyright}
\copyrightyear{2018}
\acmYear{2018}
\acmDOI{XXXXXXX.XXXXXXX}

\acmConference[Conference acronym 'XX]{Make sure to enter the correct
  conference title from your rights confirmation emai}{June 03--05,
  2018}{Woodstock, NY}

\newcommand{\ignore}[1]{}

\newcommand{\vlpa}{$\nu$-LPA}

\begin{document}

\title{Memory Efficient GPU-based Label Propagation Algorithm (LPA) for Community Detection on Large Graphs}


\author{Subhajit Sahu}
\email{subhajit.sahu@research.iiit.ac.in}
\affiliation{%
  \institution{IIIT Hyderabad}
  \streetaddress{Professor CR Rao Rd, Gachibowli}
  \city{Hyderabad}
  \state{Telangana}
  \country{India}
  \postcode{500032}
}


\settopmatter{printfolios=true}

\begin{abstract}
Community detection involves grouping nodes in a graph with dense connections within groups, than between them. We previously proposed efficient multicore (GVE-LPA) and GPU-based ($\nu$-LPA) implementations of Label Propagation Algorithm (LPA) for community detection. However, these methods incur high memory overhead due to their per-thread/per-vertex hashtables. This makes it challenging to process large graphs on shared memory systems. In this report, we introduce memory-efficient GPU-based LPA implementations, using weighted Boyer-Moore (BM) and Misra-Gries (MG) sketches. Our new implementation, $\nu$MG8-LPA, using an $8$-slot MG sketch, reduces memory usage by $98\times$ and $44\times$ compared to GVE-LPA and $\nu$-LPA, respectively. It is also $2.4\times$ faster than GVE-LPA and only $1.1\times$ slower than $\nu$-LPA, with minimal quality loss ($4.7\%$/$2.9\%$ drop compared to GVE-LPA/$\nu$-LPA).
\end{abstract}



\maketitle

\section{Introduction}
\label{sec:introduction}
Research on graph-structured data has surged due to graphs' ability to model complex, real-world interactions and relationships between entities. A key area in this field is community detection, which involves identifying clusters of vertices with stronger internal connections than those to the rest of the network \cite{com-fortunato10}. These clusters, known as communities, are "intrinsic" when based solely on the network's structure and are "disjoint" if each vertex belongs to only one group \cite{com-gregory10, coscia2011classification}. Community detection helps understand a network's structure and behavior \cite{com-fortunato10, abbe2018community}, and has a wide range of applications across various fields, including healthcare \cite{salathe2010dynamics, bechtel2005lung, haq2016community}, biological network analysis \cite{kim2009centralized, rivera2010nemo, popa2011directed}, machine learning \cite{bai2024leiden, das2011unsupervised}, urban planning \cite{wang2021network, verhetsel2022regional, chen2023deciphering}, cloud computing \cite{cao2022implementation}, social network analysis \cite{uyheng2021mainstream, blekanov2021detection, la2021uncovering, kapoor2021ll}, ecology \cite{guimera2010origin}, drug discovery \cite{ma2019comparative, udrescu2020uncovering}\ignore{, the study of human brain networks \cite{he2010graph}}, and other graph-related problems \cite{stergiou2018shortcutting, meyerhenke2017parallel, slota2020scalable, henne2015label, gottesburen2021scalable, boldi2011layered, valejo2020coarsening}.

Community detection is challenging because the number and size of communities are unknown in advance \cite{com-gregory10}. As a result, heuristic methods are often used \cite{clauset2004finding, duch2005community, reichardt2006statistical, com-raghavan07, com-blondel08, com-rosvall08, com-kloster14, com-traag19, traag2023large}\ignore{\cite{, com-xie13, com-you20, com-whang13}}. The modularity metric is commonly used to assess the quality of the detected communities \cite{com-newman06}. One widely used heuristic is the Label Propagation Algorithm (LPA), also known as RAK \cite{com-raghavan07}, a diffusion-based method known for its simplicity, speed, and scalability. Compared to the Louvain method \cite{com-blondel08}, another leading algorithm known for generating high-quality communities, LPA is $2.3 - 14\times$ faster, but tends to identify communities with $3 - 30\%$ lower quality \cite{sahu2023gvelpa}. This makes LPA particularly useful for large-scale networks where speed is more critical, and a bit of reduction in community quality is acceptable. While LPA typically scores lower in modularity, it performs well in terms of Normalized Mutual Information (NMI) against ground truth \cite{peng2014accelerating}. In our evaluation of other label-propagation based methods such as COPRA \cite{com-gregory10}, SLPA~\cite{com-xie11}, and LabelRank \cite{com-xie13}, LPA proved to be the most efficient, while identifying communities of similar quality \cite{sahu2023selecting}.

Community detection is a well-studied problem, with many efforts focused on improving algorithm performance through optimizations \cite{com-rotta11, com-waltman13, com-gach14, com-traag15, com-lu15, com-ozaki16, com-naim17, com-halappanavar17, com-ghosh18, com-traag19}\ignore{\cite{, com-shi21, com-xing14, com-berahmand18, com-sattari18, com-you20, com-liu20, com-ryu16, com-zhang21, com-you22, com-aldabobi22}} and parallelization techniques \cite{com-cheong13, com-wickramaarachchi14, com-naim17, com-halappanavar17, com-ghosh18, com-bhowmik19, com-shi21, com-bhowmick22, staudt2015engineering, soman2011fast, traag2023large}\ignore{\cite{, com-lu15, com-zeng15, com-fazlali17, com-que15, com-zeitz17, com-gheibi20}}. These studies have included work on multicore CPUs \cite{staudt2015engineering, staudt2016networkit, com-fazlali17, com-halappanavar17, qie2022isolate, huparleiden}, GPUs \cite{soman2011fast, kozawa2017gpu, ye2023large, com-naim17, kang2023cugraph}, CPU-GPU hybrids \cite{com-bhowmik19, com-mohammadi20}, multi-GPU setups \cite{com-cheong13, kang2023cugraph, chou2022batched, com-gawande22}, and multi-node systems \cite{maleki2020dhlp, ma2018psplpa, com-ghosh18, ghosh2018scalable, sattar2022scalable, huparleiden, com-bhowmick22}. The primary focus of these studies has been on reducing computation time, while memory consumption has often been a secondary concern. Yet, as network sizes continue to grow, managing memory usage is becoming increasingly crucial, especially for processing large-scale graphs in shared-memory settings. Our proposed multicore implementation of LPA, GVE-LPA \cite{sahu2023gvelpa}, offers state-of-the-art performance on shared-memory systems, but it also incurs significant memory overhead due to the reliance on per-thread hashtables. Recently, we introduced a GPU-based implementation of LPA, \vlpa{} \cite{sahu2024nulpa}, which employs a novel open-addressing per-vertex hashtable with hybrid quadratic-double probing for efficient collision resolution. However, the memory demand for hashtables scales with $O(|E|)$, where $|E|$ is the number of edges --- which can be substantial. For a graph with $4$ billion edges, \vlpa{} necessitates approximately 64 GB of GPU memory for the hashtables alone. On a $3.8$ billion-edge graph, the combined memory requirement of \vlpa{} --- including the graph storage --- surpasses the $80$ GB device memory limit of an NVIDIA A100 GPU. These limitations motivated us to explore ways to lower memory usage in community detection algorithms, even if it means sacrificing some performance.

This report presents $\nu$BM-LPA and $\nu$MG8-LPA\footnote{\url{https://github.com/puzzlef/rak-lowmem-communities-cuda}}, our memory-efficient GPU-based implementations of LPA. $\nu$BM-LPA is based on weighted Boyer-Moore majority vote, while $\nu$MG8-LPA relies on the weighted Misra-Gries heavy hitters method with $k = 8$ slots. Both implementations incorporate a Pick-Less (PL) strategy for symmetry breaking to prevent repeated label swaps. $\nu$MG8-LPA also leverages warp-level primitives for fast sketch updates and uses multiple sketches for high-degree vertices, merging them later to reduce contention\ignore{and boost performance}. Additionally, $\nu$MG8-LPA avoids rescanning the top-$k$ labels\ignore{, improving speed without compromising community detection quality}. These optimizations allow our algorithms to achieve good performance and quality of identified communities, at a significantly smaller working set size.

\section{Related work}
\label{sec:related}
Label Propagation Algorithm (LPA) is widely used in various fields, such as cross-lingual knowledge transfer for part-of-speech tagging \cite{das2011unsupervised}, 3D point cloud classification \cite{wang2013label}, sectionalizing power systems \cite{aziz2023novel}, finding connected components \cite{stergiou2018shortcutting}, graph compression \cite{boldi2011layered}, link prediction \cite{mohan2017scalable, xu2019distributed}, and graph partitioning \cite{meyerhenke2017parallel, slota2020scalable, bae2020label}. Significant work has also been conducted to improve the original LPA through various modifications \cite{farnadi2015scalable, zarei2020detecting, sattari2018spreading, zheng2018improved, zhang2017label, berahmand2018lp, el2021wlni, roghani2021pldls, li2015parallel, zhang2023large}.

Some open-source tools for LPA-based community detection include the Fast Label Propagation Algorithm (FLPA) \cite{traag2023large}, which speeds up LPA by only processing vertices with recently updated neighbors. NetworKit \cite{staudt2016networkit}, a large-scale graph analysis package with a Python interface, features a parallel LPA implementation that tracks active nodes and uses guided parallel processing.

We have proposed a high-performance multicore implementation of LPA, GVE-LPA \cite{sahu2023gvelpa}. It uses collision-free per-thread hashtables --- each hash table has a key list, a values array (sized to the number of vertices, $|V|$), and a key count. Values are stored or accumulated at indices corresponding to their keys. To prevent cache conflicts, the key count is updated independently and allocated separately in memory. This approach substantially reduces conditional branching and minimizes the number of instructions needed for inserting or accumulating entries. GVE-LPA achieves performance improvements of $139\times$ over FLPA and $40\times$ over NetworKit LPA.

Although GVE-LPA is computationally efficient, it comes with a significant memory overhead\ignore{ because it relies on per-thread full-size hashtables}. Its space complexity, not counting the input graph, is $O(T|V|)$, where $|V|$ is the number of vertices and $T$ is the number of threads used. For example, processing a graph with $200$ million vertices using 64 threads requires between $102$ and $205$ GB of memory just for the hashtables. 

Some GPU-based implementations of LPA have been introduced. Soman and Narang \cite{soman2011fast} proposed a parallel GPU algorithm for weighted LPA, while Kozawa et al. \cite{kozawa2017gpu} developed a GPU-accelerated LPA that can handle datasets too large for GPU memory. More recently, Ye et al. \cite{ye2023large} introduced GLP, a GPU-based LPA framework. However, despite the utility of LPA, there was a lack of efficient and widely available GPU-based implementation of LPA. To address this, we proposed \vlpa{}. It employs asynchronous execution, a pick-less strategy to reduce community swaps, and a novel per-vertex hashtable with hybrid quadratic-double probing for collision resolution. Running on an NVIDIA A100 GPU, it outperformed FLPA, NetworKit LPA, and GVE-LPA by $364\times$, $62\times$, and $2.6\times$, respectively. However, as mentioned earlier, the memory demand of \vlpa{}'s hashtables scale with $O(|E|)$, which is significant\ignore{, and would not be able to support processing of very large graphs}.

We now review studies\ignore{ on community detection} in the edge streaming setting, where graphs are processed as sequences of edges in a single pass. These algorithms aim to minimize runtime and memory usage for efficient graph processing. Hollocou et al. \cite{hollocou2017linear, hollocou2017streaming} introduced SCoDA, which tracks only a few integers per node by noting that edges are more likely to connect nodes within the same community. Wang et al. \cite{wang2023streaming} focused on finding local communities around query nodes by sampling neighborhoods and using an approximate conductance metric. Liakos et al. \cite{liakos2017coeus, liakos2020rapid} explored expanding seed-node sets as edges arrive without storing the full graph. Although these methods are efficient, the single-pass limit may reduce community quality compared to multi-pass algorithms.

We recently proposed to replace the per-thread hashtables in GVE-LPA with a weighted version of the Misra-Gries (MG) sketch \cite{sahu2024memory}. Our experiments showed that MG sketches with $8$ slots for LPA significantly lower memory usage the memory usage of the implementations --- while suffering only up to $1\%$ decrease in community quality, but with runtime penalties of $2.11\times$. Additionally, we presented a weighted Boyer-Moore (BM) variant for LPA which demonstrated good performance on specific graph types. In this report, we extend these methods to \vlpa{}.

\section{Preliminaries}
\label{sec:preliminaries}
Consider an undirected graph $G(V, E, w)$, where $V$ is the set of vertices, $E$ is the set of edges, and $w_{ij}$ is the weight of the edge between vertices $i$ and $j$ (with $w_{ij} = w_{ji}$). For an unweighted graph, each edge has a unit weight ($w_{ij} = 1$). The neighbors of vertex $i$ are $J_i = \{ j \mid (i, j) \in E \}$, and the weighted degree of vertex $i$ is $K_i = \sum_{j \in J_i} w_{ij}$. The graph has $N = |V|$ vertices, $M = |E|$ edges, and the total sum of edge weights is $m = \frac{1}{2} \sum_{i, j \in V} w_{ij}$.

\subsection{Community detection}
\label{sec:about-communities}

Disjoint community detection seeks to assign each vertex $i \in V$ to a community $c$ from a set $\Gamma$, via a community membership function $C: V \rightarrow \Gamma$. The set of vertices in community $c$ is denoted as $V_c$, and the community to which vertex $i$ belongs is denoted as $C_i$. For a given vertex $i$, its neighbors in community $c$ are represented as $J_{i \rightarrow c} = \{ j \ | \ j \in J_i \text{ and } C_j = c \}$, and the sum of the edge weights between $i$ and its neighbors in $c$ is $K_{i \rightarrow c} = \sum_{j \in J_{i \rightarrow c}} w_{ij}$. The total weight of edges within community $c$ is denoted by $\sigma_c = \sum_{(i, j) \in E \text{ and } C_i = C_j = c} w_{ij}$, while the total edge weight associated with community $c$ is given by $\Sigma_c = \sum_{(i, j) \in E \text{ and } C_i = c} w_{ij}$.

\subsection{Modularity}
\label{sec:about-modularity}

Modularity is a measure of the quality of communities identified by community detection algorithms. It calculates the difference between the actual fraction of edges within communities and the expected fraction if edges were randomly assigned, with values ranging from $[-0.5, 1]$. Higher values indicate stronger community structure. The modularity $Q$ is computed using Equation \ref{eq:modularity} and involves the Kronecker delta function ($\delta(x,y)$), which equals 1 when $x = y$ and 0 otherwise. Additionally, the \textit{delta modularity} for moving vertex $i$ from community $d$ to community $c$ is denoted as $\Delta Q_{i: d \rightarrow c}$, calculated with Equation \ref{eq:delta-modularity}.

\begin{equation}
\label{eq:modularity}
  Q
  = \frac{1}{2m} \sum_{(i, j) \in E} \left[w_{ij} - \frac{K_i K_j}{2m}\right] \delta(C_i, C_j)
  = \sum_{c \in \Gamma} \left[\frac{\sigma_c}{2m} - \left(\frac{\Sigma_c}{2m}\right)^2\right]
\end{equation}

\begin{align}
\begin{split}
\label{eq:delta-modularity}
  &\Delta Q_{i: d \rightarrow c} = \Delta Q_{i: d \rightarrow i} + \Delta Q_{i: i \rightarrow c} \\
  &= \left[ \frac{\sigma_d - 2K_{i \rightarrow d}}{2m} - \left(\frac{\Sigma_d - K_i}{2m}\right)^2 \right] + \left[ 0 - \left(\frac{K_i}{2m}\right)^2 \right] - \left[ \frac{\sigma_d}{2m} - \left(\frac{\Sigma_d}{2m}\right)^2 \right] \\
  &+ \left[ \frac{\sigma_c + 2K_{i \rightarrow c}}{2m} - \left(\frac{\Sigma_c + K_i}{2m}\right)^2 \right] - \left[ \frac{\sigma_c}{2m} - \left(\frac{\Sigma_c}{2m}\right)^2 \right] - \left[ 0 - \left(\frac{K_i}{2m}\right)^2 \right] \\
  &= \frac{1}{m} (K_{i \rightarrow c} - K_{i \rightarrow d}) - \frac{K_i}{2m^2} (K_i + \Sigma_c - \Sigma_d)
\end{split}
\end{align}

\subsection{Label Propagation Algorithm (LPA)}
\label{sec:about-rak}

LPA \cite{com-raghavan07} is a fast, scalable diffusion-based method for detecting moderate-quality communities in large networks, outperforming the Louvain method \cite{com-blondel08} in terms of simplicity and speed. Initially, each vertex has a unique label (community ID). In each iteration, vertices update their labels by adopting the one with the highest total connecting weight, as described in Equation \ref{eq:rak}. This process continues until a consensus is reached, forming communities. The algorithm stops when at least $1-\tau$ of the vertices (where $\tau$ is a tolerance parameter) keep their labels unchanged. LPA has a time complexity of $O(L |E|)$ and space complexity of $O(|V| + |E|)$, where $L$ is the number of iterations \cite{com-raghavan07}.

\begin{equation}
\label{eq:rak}
  C_i =\ \underset{c\ \in \ \Gamma}{\arg\max} { \sum_{j \in J_i\ |\ C_j = c} w_{ij} }
\end{equation}

\subsection{Boyer-Moore (BM) majority vote algorithm}
\label{sec:about-bm}

The Boyer-Moore majority vote algorithm efficiently identifies the majority element in a sequence, which is an element that appears more than $n/2$ times in a list of $n$ elements. Developed by Boyer and Moore \cite{boyer1991mjrty} in 1981, the algorithm tracks a candidate and a vote count. Initially, the candidate is set, and the count starts at zero. As the list is traversed, if the count is zero, the current element becomes the new candidate with a count of one. If the current element matches the candidate, the count increases; if it differs, the count decreases. At the end, the candidate holds the potential majority element. It runs in $O(n)$ time and uses $O(1)$ space.

\subsection{Misra-Gries (MG) heavy hitters algorithm}
\label{sec:about-mg}

The Misra-Gries (MG) heavy hitters algorithm, introduced in 1982 by Misra and Gries \cite{misra1982finding}, extends the Boyer-Moore majority algorithm to identify elements that occur more than $\frac{n}{k+1}$ times, where $n$ is the total number of elements and $k+1$ is a user-defined threshold. The algorithm uses up to $k$ counters to track candidate elements and their counts. If a candidate appears, its counter is incremented; if not, a new counter is created if space allows. When all counters are full, each counter is decremented, and elements with zero counts are removed. After processing, the remaining candidates are likely to exceed the $\frac{n}{k+1}$ threshold, though a verification step is needed. The algorithm runs in $O(n)$ time and uses $O(k)$ space, making it efficient for limited-resource environments.

\subsection{Fundamentals of a GPU}
\label{sec:about-gpu}

The core unit of NVIDIA GPUs is the Streaming Multiprocessor (SM), which contains multiple CUDA cores for parallel processing. Each SM also has shared memory, registers, and specialized function units. The number of SMs varies by GPU model, and each operates independently. The memory hierarchy includes global memory (largest but slowest), shared memory (low-latency, used by threads within an SM), and local memory (private storage for threads when registers are full). Threads on a GPU are organized into warps (32 threads executing together), thread blocks (groups of threads on the same SM), and grids (collections of thread blocks). Warps execute in lockstep, and SMs schedule warps alternately if threads stall. Threads within a block communicate via shared memory, while blocks in a grid exchange data through global memory\ignore{, which is slower than shared memory but allows inter-block communication}.

\subsection{Warp-level primitives}
\label{sec:about-warp-primitives}

NVIDIA GPUs offer warp-level primitives that allow operations to be executed in parallel across all threads in a warp, enabling efficient communication. A key function is the \textit{warp vote} operation, which performs logical operations across all threads in a warp. These include \texttt{\_\_all\_sync()}, \texttt{\_\_any\_sync()}, and \texttt{\_\_ballot\_sync()}. \texttt{\_\_al} \texttt{l\_sync()} checks if all threads satisfy a condition, \texttt{\_\_any\_sync()} checks if at least one does, and \texttt{\_\_ballot\_sync()} collects each thread's boolean result in a 32-bit integer.

\section{Approach}
\label{sec:approach}
\begin{figure}[hbtp]
  \centering
  \subfigure{
    \label{fig:about-hashtable--all}
    \includegraphics[width=0.86\linewidth]{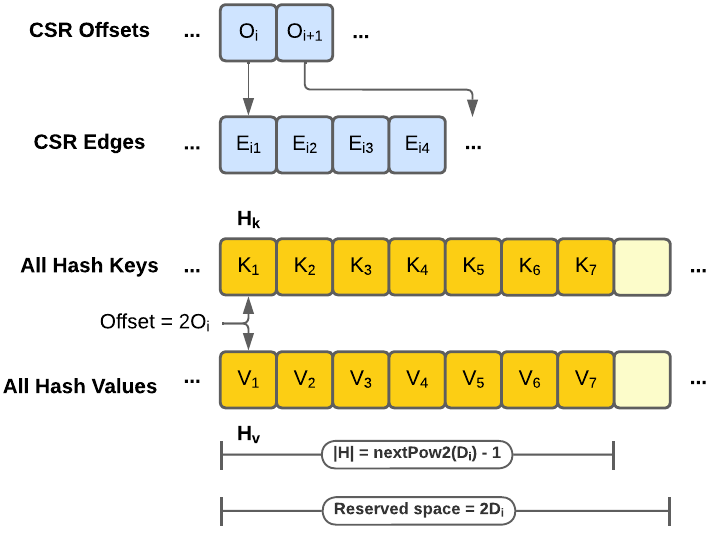}
  } \\[-2ex]
  \caption{Illustration of per-vertex open-addressing hashtables in $\nu$-LPA \cite{sahu2024nulpa}. Each vertex $i$ has a hashtable $H$ with a key array $H_k$ and a value array $H_v$. Memory for all hash key and value arrays is allocated together. The offset for vertex $i$'s hashtable is $2O_i$, where $O_i$ is its CSR offset. The total memory for the hashtable is $2D_i$, where $D_i$ is the vertex's degree. The hashtable’s capacity is $nextPow2(D_i) - 1$.}
  \label{fig:about-hashtable}
\end{figure}

We recently introduced $\nu$-LPA \cite{sahu2024nulpa}, a GPU-optimized version of LPA based on GVE-LPA \cite{sahu2023gvelpa}. It uses per-vertex open-addressing hashtables, as shown in Figure \ref{fig:about-hashtable}, where the size of the hashtables is proportional to each vertex's degree. $\nu$-LPA does this because allocating large fixed-size hashtables per thread, similar to GVE-LPA, is infeasible on GPUs, which support massive parallelism but have a limited memory. However, as brought up earlier, $\nu$-LPA still requires significant memory, with a space complexity of $O(|E|)$, where $|E|$ is the number of edges. For instance, processing a graph with $4$ billion edges requires $64$ GB of GPU memory for the hashtables alone. Memory demand can escalate rapidly, as graphs grow. This highlights the need to explore ways of reducing the memory footprint of the hashtables, even at the cost of some performance.

Note that in every iteration of LPA, each vertex $i \in V$ iterates over its neighbors $J_i$, excluding itself, and calculates the total edge weight $K_{i \rightarrow c}$ for each unique label $c \in \Gamma_i$ among its neighbors. These weights are stored in a hashtable. The label $c^*$ with the highest weight $K_{ i \rightarrow c^*}$ is then selected as the new label for vertex $i$.

To reduce the memory usage of LPA, we focus on a ``sketch'' of neighboring community labels rather than storing a fully populated map. Instead of keeping all labels $c \in \Gamma_i$ for each vertex $i$ and their associated linking weights $K_{i \rightarrow c}$, we only track labels with a linking weight greater than $\frac{K_i}{k+1}$, where $K_i$ is the weighted degree of $i$, and $k$ is a user-defined parameter. The intuition is that the label with the highest linking weight, $c^*$, will likely be among the $k$ most significant labels. To achieve this, we use a weighted version of the Misra-Gries (MG) heavy-hitter algorithm \cite{misra1982finding} with $k$ slots. Instead of counting occurrences of neighboring community labels, we accumulate the edge weights between vertex $i$ and its neighbors, grouped by community label. We then identify up to $k$ candidate labels. Not all of these labels will necessarily have a linking weight above $\frac{K_i}{k+1}$, so some entries may correspond to non-majority labels or remain empty if there are fewer than $k$ labels. In a second scan, we may calculate the total linking weight between $i$ and the candidate labels, and select the label $c^\#$ with the highest weight. While $c^\#$ may differ from the highest weight label $c^*$, the two are expected to align in most cases when $k$ is appropriately chosen \cite{sahu2024memory}.

Furthermore, we investigate reducing the memory usage of our GPU implementation of LPA by employing a weighted variant of the Boyer-Moore (BM) majority vote algorithm \cite{boyer1991mjrty}. This approach represents a minimal case of the weighted MG algorithm with $k = 1$, where only a single majority candidate label is tracked \cite{sahu2024memory}.

We now present the design of our GPU-based implementation of LPA, called $\nu$MG-LPA, which uses $k$ slots to maintain a sketch or summary of neighboring community labels for each vertex, leveraging a weighted version of the Misra-Gries (MG) heavy hitters algorithm. Building on this, $\nu$BM-LPA --- our GPU-based LPA implementation based on the weighted Boyer-Moore (BM) majority vote algorithm --- is described later, in Section \ref{sec:bmlpa}.

\subsection{Design of MG Sketch}
\label{sec:mg-sketch-design}

We utilize $k$ slots in our weighted MG sketch $S$, which is composed of two distinct arrays: $S_k$ and $S_v$. The $S_k$ array holds the candidate community labels for the current vertex being processed, while the $S_v$ array stores the corresponding sketch weights. Each slot in the sketch is identified by an index $s$, where $s < k$. A slot $s$ is deemed empty if its associated weight is zero, meaning $S_v[s] = 0$.

We now explain how a key-value pair $(c, w)$ is accumulated into the MG sketch $S$. The process begins by checking whether the community label $c$ already exists as a candidate label in the sketch, i.e., if $S_k[s] = c$ for some $s$. If $c$ is found, the associated weight of the slot is incremented by the edge weight $w$, updating $S_v[s] \gets S_v[s] + w$. If $c$ does not exist as a candidate label, an attempt is made to populate a free slot $s_\phi$ (where $S_v[s_\phi] = 0$) with the key-value pair by setting $S_k[s_\phi] = c$ and $S_v[s_\phi] = w$. If no free slot is available (i.e., all slots are occupied, with $S_v[s] \neq 0$ for all $s$), the associated weights of all slots in the sketch are decremented by the edge weight $w$, applying $S_v[s] \gets S_v[s] - w$ for all $s$. This decrement ensures that less frequent labels are gradually removed, freeing up space for labels that may become more frequent.

To efficiently accumulate key-value pairs into the MG sketch using threads, we avoid assigning the task of updating/managing an MG sketch to a single thread, as this would not leverage the parallelism of GPUs and would introduce significant warp divergence. Instead, we assign at least one thread group $g$ to manage each MG sketch, where each slot $s$ in the sketch is exclusively managed a unique thread $t$ within the thread group, in parallel. Cooperative groups \cite{harris2017cooperative} are employed to partition the threads in a thread block (see Section \ref{sec:about-gpu} for more details) into smaller thread groups. This partitioning is essential because cooperative groups allow for the creation of thread groups smaller than a warp (32 threads), which would otherwise be restricted to MG sketches with at least 32 slots. We use \texttt{tiled\_partition()} to partition threads in a thread block, with the size of each cooperative group / thread group, being fixed at compile time to $k$. Additionally, given that each MG sketch is relatively small but subjected to many updates, we store the MG sketch for each thread group in shared memory, which acts as a user-managed cache, providing significantly higher memory bandwidth than global memory (which is external).

When accumulating a key-value pair $(c, w)$ into the MG sketch, two communication points between threads in the thread group are required: one to check if a community label $c$ already exists as a candidate label in the sketch, and another to find a free slot $s_\phi$ to store $(c, w)$ if $c$ is not already present. This intra-group communication can be handled using shared memory variables. We would like to note that, with a suitable choice of representative values, it is possible to get away with a single shared memory variable, which we refer to as $has$. In particular, we initialize $has$ to $-1$ to indicate that $c$ does not exist as a candidate label, and set to $0$ if a matching candidate label is found. If no candidate label exists (i.e., $has = -1$), each thread $t$ in the thread group\ignore{performing operations on slot $s$} executes an atomic max operation on $has$ to determine the last free slot in the sketch, which can then be populated with $(c, w)$. If no free slot is found, $has$ remains set to $-1$. Afterward, all threads decrement their respective slots by $w$. Note that we must synchronize all threads in the group, before accessing $has$, to ensure proper conditional branching.

An MG sketch may however be shared between multiple thread groups, i.e., each slot in the sketch may no longer be updated by a single thread exclusively but rather operated upon by multiple threads (one per thread group). For such shared sketches, atomic operations must be employed when updating them. Specifically, atomic addition (\texttt{atomicAdd()}) is required to increment the associated weight $S_v[s]$ of a matching candidate label at slot $s$ by edge weight $w$ when accumulating a key-value pair $(c, w)$, and to decrement the weights of all slots in the sketch by $w$ when no free slots exist. In addition, atomic compare-and-swap (\texttt{atomicCAS()}) must be used when populating a free slot $s_\phi$ with $(c, w)$. Since multiple threads may attempt to populate the same free slot, the \texttt{atomicCAS()} operation can fail, necessitating a retry loop to find another available free slot. Furthermore, shared variables like $has$, which are used for communication between threads, should be updated atomically, although atomicity is not always required in certain cases, such as with a boolean-like variable that only changes in one direction. This applies to the shared variable $has$, when it is used to identify whether a matching candidate label exists in the sketch for the key-value pair being accumulated.

\begin{figure}[hbtp]
  \centering
  \subfigure{
    \label{fig:optslots--runtime}
    \includegraphics[width=0.98\linewidth]{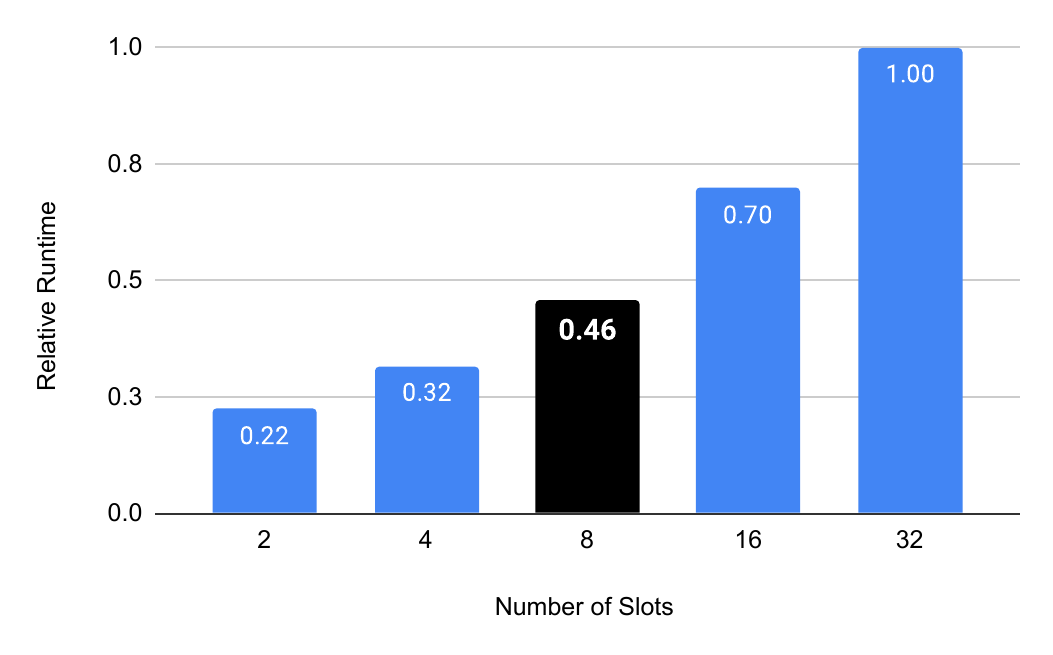}
  }
  \subfigure{
    \label{fig:optslots--modularity}
    \includegraphics[width=0.98\linewidth]{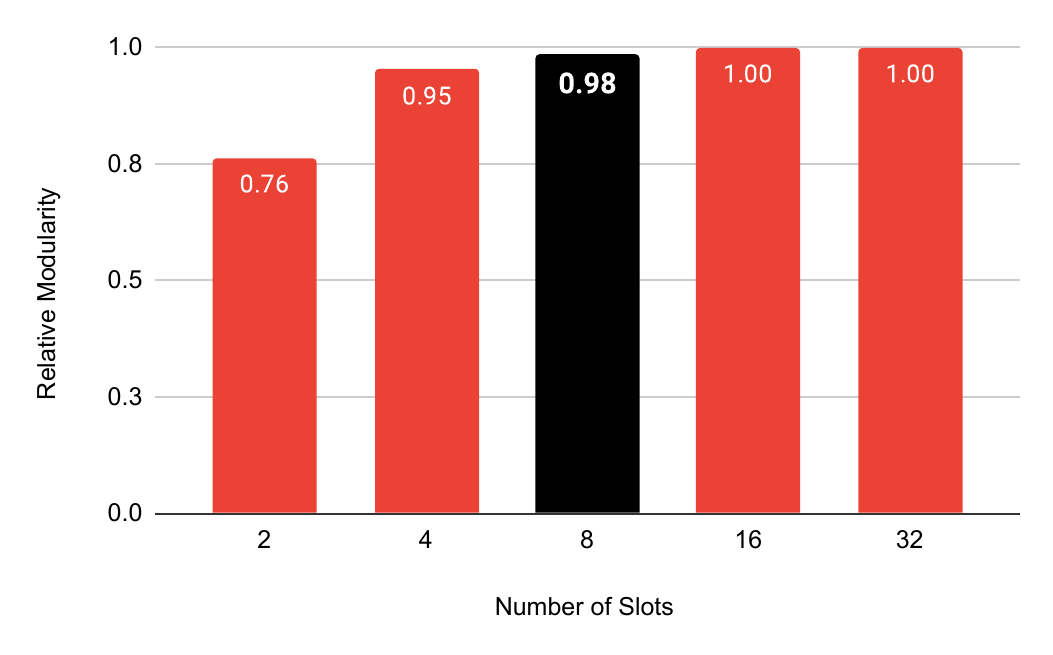}
  } \\[-2ex]
  \caption{Relative runtime and Modularity of obtained communities using $\nu$MG-LPA, with varying number of slots $k$ in the Misra-Gries (MG) sketch, ranging from $2$ to $32$.}
  \label{fig:optslots}
\end{figure}

We now determine a suitable value for $k$, the number of slots in each MG sketch, for our GPU implementation of LPA. A larger value of $k$ is expected to improve community quality, as it increased the likelihood of identifying the most weighted label for a given vertex. However, increasing $k$ also requires a larger number of threads per thread group, which reduces the number of vertices processed per unit time. Additionally, a higher $k$ leads to increased communication and synchronization costs between threads, as well as lower occupancy of Symmetric Multiprocessors (SMs). We experiment with $k$ values ranging from $2$ to $32$ (in powers of $2$) on large, real-world graphs (see Table \ref{tab:dataset}), ensuring each graph is undirected and weighted, with edge weights set to $1$. Figure \ref{fig:optslots} illustrates the relative runtime and modularity of communities for varying values of $k$. The results show that using MG sketches with $k = 8$ slots runs $2.2\times$ faster than those with $k = 32$, while the community quality only decreases by $1.6\%$, striking a balance between runtime and community quality. Therefore, we select $k = 8$ slots for each sketch.

Since CUDA 9, NVIDIA has introduced warp-level primitives \cite{lin2018using}, which enable threads within a warp to directly exchange data, perform collective operations, and coordinate their execution without relying on shared memory or synchronization primitives like barriers. These include warp-level vote functions, such as \texttt{\_\_all\_sync()} and \texttt{\_\_ballot\_sync()}, which, as detailed in Section \ref{sec:about-warp-primitives}, allow threads in a warp to check whether all threads in the warp (or a subset, depending on selected thread flags) satisfy a condition, or to collect each thread's boolean result in a 32-bit integer. Additionally, cooperative groups \cite{harris2017cooperative} provide a simpler API for these warp-level primitives. To leverage this, we replace the use of the shared memory variable $has$ as follows: Given a key-value pair $(c, w)$ to accumulate into the sketch $S$, each thread $t$ in the thread group $g$ (managing the sketch) checks if its respective slot $s$ in the sketch has $c$ as the candidate label, i.e., $S_k[s] = c$. A call to the \texttt{g.ballot()} function is used to check if any thread in $g$ successfully found $c$ as the candidate label in their respective slots. A similar procedure is followed to find a free slot, where threads collaborate using the ballot function, and \texttt{\_\_ffs()} (which finds the first set bit) is used to determine the first available slot in the sketch. We also use \texttt{g.all()} to check if all threads in $g$ failed to populate a free slot, in which case, the weights of all slots in the sketch are decremented by the edge weight $w$. Note that using warp-level primitives limits the number of slots $k$ in our MG sketch to $32$, the size of a warp. However, this is not problematic, as we have already identified $k = 8$ slots to be suitable for our algorithm.

\begin{figure}[hbtp]
  \centering
  \subfigure{
    \label{fig:optwarp--runtime}
    \includegraphics[width=0.98\linewidth]{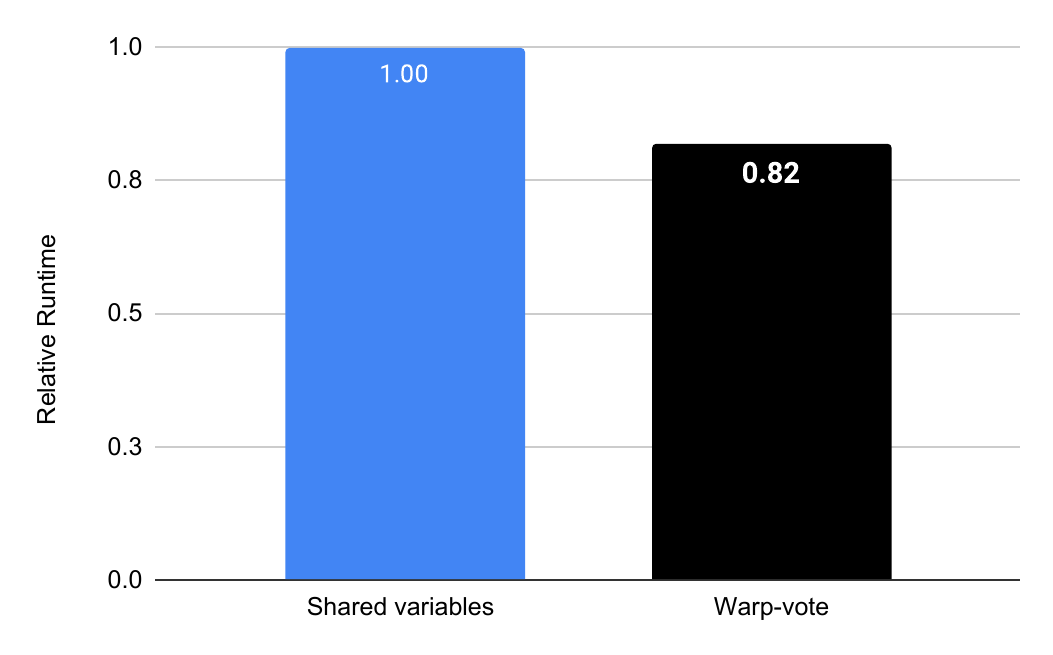}
  } \\[-2ex]
  \caption{Relative Runtime of \textit{Shared variables} and \textit{Warp-vote} approaches for populating weighted Misra-Gries (MG) sketches from the neighborhood of each vertex.\ignore{See Section \ref{sec:mg-sketch-design} for details of each approach.}}
  \label{fig:optwarp}
\end{figure}

Based on the above discussion, we analyze two variations of our GPU-based implementation of LPA: the \textit{Shared variables} approach and the \textit{Warp-vote} approach. In the \textit{Shared variables} approach, each neighbor of a vertex, along with its associated edge weight, is accumulated into the MG sketch using shared memory variables for intra-group communication. In contrast, the \textit{Warp-vote} approach uses warp-level voting functions, instead of shared memory, to support thread cooperation in populating the sketch. We conduct experiments on large graphs (shown in Table \ref{tab:dataset}), ensuring that each graph is undirected and weighted, as earlier. Figure \ref{fig:optwarp} illustrates the relative runtime of the \textit{Shared variables} and \textit{Warp-vote} approaches for populating/accumulating MG sketches. As the figure shows, with the \textit{Warp-vote} approach is $1.2\times$ faster than \textit{Shared variables} approach. Both approaches result in communities with the same modularity, and hence, is not shown in the figure. Accordingly, we use the \textit{Warp-vote} approach to populate the sketch.

\subsection{Organization of Thread Groups}
\label{sec:mg-thread-groups}

We now discuss how we organize thread groups to update the labels of vertices in the input graph. A key step in this process is obtaining an MG sketch of each vertex's neighborhood, which helps identify the candidate label, $c^\#$, with the highest linking weight. This label is then selected as the updated label for the vertex. A simple approach to achieve this would be to assign a thread group to each vertex, where the group is responsible for managing/updating the sketch. However, many real-world graphs follow a power-law degree distribution, where a small set of vertices have high degrees while the majority have low degrees. If high-degree vertices are processed by a single thread group, it would likely lead to significant load imbalance. To address this, a better approach is to assign multiple thread groups to high-degree vertices, ideally in proportion to their degree. This improves parallelism but can introduce increased contention and repeated retries for occupying free slots in the sketch.

However, to keep things simple, and avoid the overhead of multiple kernel calls (each corresponding to a specific number of thread groups per vertex), we instead partition the vertices into two sets: high-degree and low-degree vertices --- based on a degree threshold $D_H$, where vertices with degree $\geq D_H$ are classified as high-degree. We then process low-degree vertices using a group-per-vertex kernel, where each vertex is handled by a single thread group; and high-degree vertices using a block-per-vertex kernel, where each vertex is processed by one thread block, consisting of $R_H$ thread groups. For high-degree vertices processed by the block-per-vertex kernel, the MG sketch is ``shared'', i.e., each slot in the sketch is updated by multiple threads (one in each thread group), necessitating the use of atomic operations and retry loops when updating, as detailed in Section \ref{sec:mg-sketch-design}. In contrast, for low-degree vertices processed by the group-per-vertex kernel, the MG sketch is not shared --- eliminating the need for atomic operations or retry loops\ignore{, and making the process more efficient}.

We now discuss a few additional operations needed to update the label for each vertex. In order to find the candidate label $c^\#$ with the most linking weight for each vertex $i \in V$, once the MG sketch $S$ has been populated, we do a second scan to calculate the total linking weight between $i$ and the candidate labels, and then perform a pairwise max block-reduce on the sketch labels array $S_k$ and weights array $S_v$. If $c^\#$ differs from the current label of the vertex, $C[i]$, we update $C[i]$ to $c^\#$ and, in parallel, mark the neighbors of $i$ as unprocessed. Additionally, one thread in a thread group/thread block tracks the number of label updates observed, denoted as $\Delta N_G$, which is then atomically added to $\Delta N$, a global variable that counts the number of changed vertices in the current iteration. To ensure correctness and avoid race conditions when threads within a block share data or depend on each other for computations, appropriate synchronization barriers are employed throughout the kernel.

To optimize the parameters of our algorithm, which include the degree threshold $D_H$ for high-degree vertices (processed by the block-per-vertex kernel), the number of thread groups $R_H$ used per vertex in the block-per-vertex kernel, and the kernel launch configurations for both the group- and block-per-vertex kernels, we performed manual gradient descent. This involved iteratively adjusting each parameter slightly and observing its impact on runtime, with random cycling through parameters until further optimization was no longer achievable. After numerous adjustments, we determined optimal values as follows: a degree threshold $D_H$ of $128$, a per-vertex thread group count $R_H$ of $32$ for the block-per-vertex kernel, and kernel launch configurations of $32$ threads per block for the group-per-vertex kernel and $256$ threads per block for the block-per-vertex kernel. Recognizing the potential for settling in a local minimum, we aim to explore auto-tuning of the kernels in the future, which is especially important for AMD GPUs \cite{lurati2024bringing}.

\subsection{Consolidation of Sketches}
\label{sec:mg-sketch-merge}

As discussed earlier, we assign multiple thread groups to high-degree vertices in the block-per-vertex kernel to process multiple edges and update the MG sketch in parallel. However, shared sketches can experience significant contention, especially when the number of cooperating thread groups, $R_H$, is large --- such as the value $R_H = 32$ in our case. Moreover, with a large $R_H$, operations like finding a free slot may frequently fail as threads compete to fill any remaining free slots in the sketch, leading to increased retries and potential performance degradation due to warp divergence, as the size of each thread group ($k = 8$) is smaller than the warp size.

However, Misra-Gries (MG) sketches, also known as MG summaries, are mergeable \cite{agarwal2013mergeable}. Therefore, to address the above issue, we consider the use of separate sketches $S[g]$ for each thread group $g$ processing a vertex (covering a subset of its neighbors). We also refer to these separate sketches as \textit{partial} sketches, as they represent the sketch of a subset of neighbors of each vertex. Once all edges of the vertex are processed, and all the partial sketches are populated, they are merged --- in the block-per-vertex kernel. Using separate sketches per thread group implies that we no longer need atomic operations and retry loops when operating on such sketches. However, this does involve additional work of merging the independent sketches, and could lower the occupancy of SMs due to the increased\ignore{amount of} shared memory needed per thread block.

To merge the independent partial sketches $S[g]$ from each thread group $g$ into a single consolidated sketch, all thread groups except the first ($g \neq 0$) work in parallel to merge their private sketches $S[g]$ into the sketch of the first thread group, $S[0]$. Specifically, each thread group $g$ iteratively accumulates non-empty slots, which contain candidate labels and their associated weights, from its own sketch $S[g]$ into $S[0]$ until all its slots are processed. During this process, each thread within $g$ is assigned to operate on a slot in $S[0]$ in a shared manner using atomic operations and retry loops.

\begin{figure}[hbtp]
  \centering
  \subfigure{
    \label{fig:optmerge--runtime}
    \includegraphics[width=0.98\linewidth]{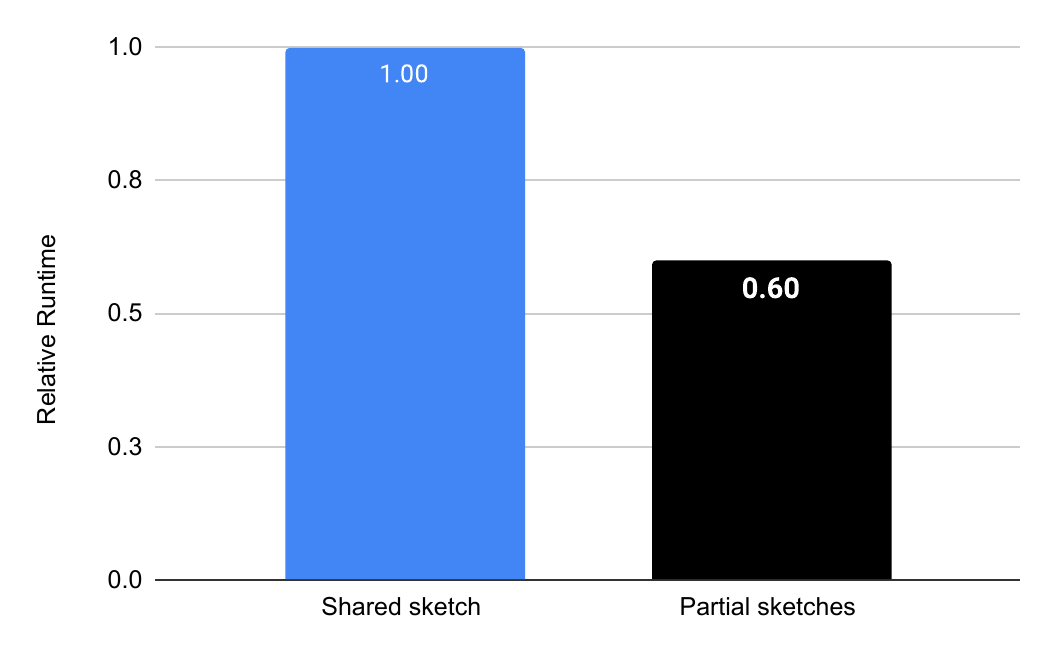}
  } \\[-2ex]
  \caption{Relative Runtime of \textit{Shared sketch} and \textit{Partial sketches} approaches for populating weighted Misra-Gries (MG) sketches from the neighborhood of each vertex.\ignore{See Section \ref{sec:mg-sketch-design} for details of each approach.}}
  \label{fig:optmerge}
\end{figure}

The merging step discussed above introduces some contention, but since the work involved is minimal, we believe the cost is negligible. To confirm this, we conduct an experiment comparing the performance of two approaches: the \textit{Shared sketch} approach, which uses warp-level voting functions to populate a single shared sketch (in the block-per-vertex kernel), and the \textit{Partial sketches} approach, which also utilizes warp-level voting functions but employs separate sketches for each thread group processing a vertex, where each group populates its own sketch based on the neighbors and associated edge weights it observes, and later merges these into a consolidated sketch in parallel. It is important to note that the group-per-vertex kernel is identical for both approaches. The experiment was conducted on the graphs from Table \ref{tab:dataset}, ensuring that each graph was undirected and weighted, with a weight of $1$ for each edge. Figure \ref{fig:optmerge} presents the mean relative runtime of the \textit{Shared sketch} and \textit{Partial sketches} approaches --- showing that the \textit{Partial sketches} approach is $1.7\times$ faster than the \textit{Shared sketch} approach. Since both approaches yield communities with the same modularity, this is not shown in the figure. As a result, we opt to use the \textit{Partial sketches} approach for the block-per-vertex kernel.

\subsection{A Single Scan is Sufficient}
\label{sec:mg-rescan}

Note that the $k$ candidate labels we obtain for a vertex $i$ in an MG sketch will include labels with a linking weight greater than $\frac{K_i}{k+1}$, where $K_i$ is the weighted degree of $i$. However, not all of these labels will necessarily exceed this threshold, i.e., some entries may correspond to non-majority labels, or remain empty, if there are fewer than $k$ labels. A second scan is then performed over the neighbors of vertex $i$ to compute the total linking weight between $i$ and the candidate labels, selecting the label $c^\#$ with the highest linking weight. This second scan involves first clearing the associated weights in the consolidated sketch and then accumulating the total linking weights for each candidate label by adding the edge weight $w$ of each neighbor with label $c$ into the corresponding slot in the sketch. This accumulation process adds additional computational cost. Furthermore, in the block-per-vertex approach, $R_H$ thread groups process $R_H$ edges of the vertex in parallel within the shared sketch, leading to contention between thread groups.

\begin{figure}[hbtp]
  \centering
  \subfigure{
    \label{fig:optrescan--runtime}
    \includegraphics[width=0.98\linewidth]{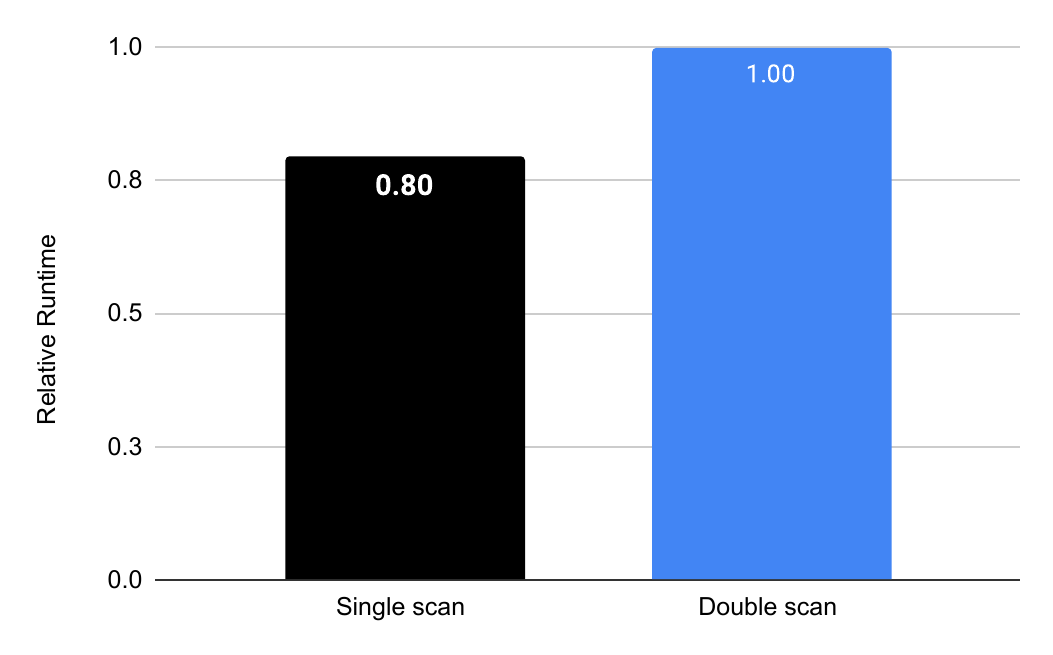}
  } \\[-2ex]
  \caption{Relative Runtime of \textit{Single scan} vs. \textit{Double scan} approaches for selecting the updated label of each vertex\ignore{in $\nu$MG-LPA}.}
  \label{fig:optrescan}
\end{figure}

However, it is likely that the most weighted candidate label in the MG sketch, which we refer to as $c^@$, will align with the label $c^\#$ that has the highest linking weight after a second scan. This eliminates the need to calculate the total linking weight $K_{i \rightarrow c}$ between the current vertex $i$ and each of its $k$-majority communities. To test this theory, we conduct an experiment comparing the performance of two approaches: the \textit{Single scan} approach, where we select $c^@$, the most weighted candidate label in the MG sketch, as the new label for the vertex, and the \textit{Double scan} approach, where we perform a second scan on the vertex's edges to calculate the total linking weight between $i$ and the candidate labels in the sketch, and then select the label $c^\#$ with the highest linking weight. The experiment is conducted on the graphs from Table \ref{tab:dataset}, ensuring that each graph is undirected and weighted. Figure \ref{fig:optrescan} illustrates the mean relative runtime of the \textit{Single scan} and \textit{Double scan} approaches. As the results show, the \textit{Single scan} approach is $1.3\times$ faster than the \textit{Double scan} approach. Both approaches yield nearly identical modularity values for the resulting communities, which is why modularity is not shown in the figure. Based on these results, we adopt the \textit{Single scan} approach to select the new label for each vertex.

\subsection{Mitigating Community Swaps}
\label{sec:mg-pickless}

Note that GPU-based LPA can fail to converge due to vertices getting stuck in cycles of community label swaps. This can happen when two interconnected vertices keep adopting each other's labels, especially in symmetrical situations where vertices are equally connected to each other's communities. Such swaps are more likely because GPUs execute in lockstep, and symmetrical vertices may end up repeatedly swapping labels, preventing convergence. Therefore, symmetry-breaking techniques are essential \cite{sahu2024nulpa}.

In our previous work \cite{sahu2024nulpa}, we introduced the Pick-Less (PL) approach to address this issue, where a vertex can only switch to a lower community ID, preventing community swaps. However, using PL too frequently can reduce the algorithm's ability to identify high-quality communities. We found that applying PL every $\rho = 4$ iterations, starting from the first iteration, results in the highest modularity communities. However, after further testing, we found that a $\rho$ value of $8$ is slightly more effective (in fact, it seems that applying PL in the first iteration resolves most community swap problems). However, conservatively, we use a value of $\rho = 8$\ignore{for $\nu$MG-LPA}.

\subsection{Our Memory Efficient GPU-based LPA employing Misra-Gries (MG) Sketch}
\label{sec:mglpa}

\begin{figure}[hbtp]
  \centering
  \subfigure[Group-per-vertex kernel of $\nu$MG-LPA]{
    \label{fig:about-move--low}
    \includegraphics[width=0.66\linewidth]{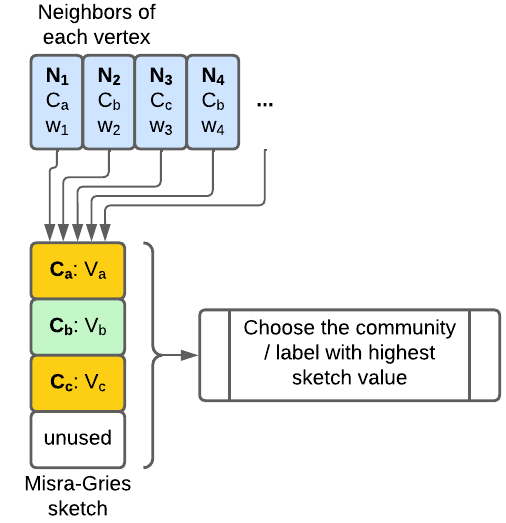}
  }
  \subfigure[Block-per-vertex kernel of $\nu$MG-LPA]{
    \label{fig:about-move--high}
    \includegraphics[width=0.66\linewidth]{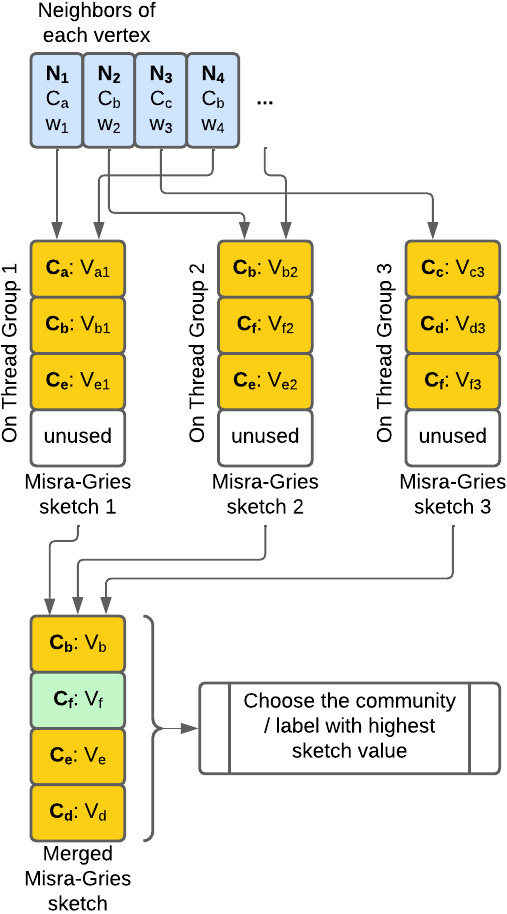}
  } \\[-2ex]
  \caption{Illustration of how $\nu$MG-LPA selects the best candidate community label for each vertex, with the group-per-vertex kernel shown in (a), and the block-per-vertex kernel shown in (b). Here, the number of slots in each MG sketch is assumed to be $k = 4$, and in the block-per-vertex approach, the number of thread groups processing the vertex is assumed to be $3$. In the figure, $N_*$ represents the neighbors of a vertex, $C_*$ represents the community labels of those neighbors, and $w_*$ represents the edge weights associated with each neighbor. Additionally, each slot in the MG sketch is associated with labels/keys $C_*$ and values/weights $V_*$.}
  \label{fig:about-move}
\end{figure}

The optimizations discussed above significantly reduce the memory usage of our weighted Misra-Gries (MG) based GPU implementation of LPA, $\nu$MG-LPA, while maintaining competitive performance in terms of runtime and community quality (modularity), when compared to $\nu$-LPA \cite{sahu2024nulpa}. A high-level overview of $\nu$MG-LPA is shown in Figure \ref{fig:about-move}, which demonstrates how $\nu$MG-LPA selects the best candidate community label for each vertex, comparing the group-per-vertex kernel in Figure \ref{fig:about-move--low} and the block-per-vertex kernel in Figure \ref{fig:about-move--high}. In both cases, each MG sketch contains $k = 4$ slots, and in the block-per-vertex kernel, each vertex is processed by $3$ thread groups (as an example). In the group-per-vertex kernel, a single thread group is assigned to each vertex, which populates the sketch in parallel using $k$ threads, and the vertex's new label is the one with the highest weighted sketch value. In the block-per-vertex kernel, multiple thread groups (in this example, $3$) are assigned to each vertex. Each thread group populates its own private sketch based on a subset of neighbors, with each group using $k$ threads, and then the separate sketches are merged into a single consolidated sketch in parallel. The vertex's new label is then chosen based on the highest weighted candidate label in this merged sketch.

Since the MG sketches of $\nu$MG-LPA are of fixed size and reside on the shared memory of the GPU, the space complexity of $\nu$MG-LPA is $O(|V|)$, excluding the input graph --- in contrast to $\nu$-LPA's space complexity of $O(|E|)$. Both algorithms have the same time complexity of $O(K|E|)$, where $K$ represents the number of LPA iterations performed. The pseudocode for $\nu$MG-LPA is presented in Algorithm \ref{alg:rakmg}, while the pseudocode for populating the MG sketches is given in Algorithm \ref{alg:sketch}, with detailed explanations in Sections \ref{sec:explain-rakmg} and \ref{sec:explain-sketch}, respectively. Note that we also refer to our algorithm as $\nu$MG8-LPA, since we use $k = 8$ slots for the MG sketches.

\subsection{Our Memory Efficient GPU-based LPA employing Boyer-Moore (BM) Algorithm}
\label{sec:bmlpa}

We now discuss the design of $\nu$BM-LPA, our GPU-based implementation of LPA based on a weighted version of the Boyer-Moore (BM) majority vote algorithm. The algorithm processes a key-value pair $(c, w)$ by first checking if the community label $c$ matches the current majority weighted label $c^\#$. If $c^\# = c$, the associated majority weight $w^\#$ is incremented by $w$. If $c^\# \neq c$, it checks whether $w^\# > w$; if so, $w^\#$ is decremented by $w$; otherwise, both $c^\#$ and $w^\#$ are updated to $c$ and $w$, respectively. For load balancing, as in $\nu$MG-LPA, vertices in the input graph are partitioned into low- and high-degree sets. However, unlike $\nu$MG-LPA, low-degree vertices are processed with a thread-per-vertex kernel, as the update can be handled by a single thread. High-degree vertices, on the other hand, are processed using a block-per-vertex kernel, where a thread block subdivides processing of the vertex's edges among multiple threads, with each thread maintaining its own $c^\#$ and $w^\#$ based on the subset of edges it observes. After all edges are processed, the threads collaborate in a pair max block-reduce operation to determine the majority $c^\#$ and $w^\#$ across all threads. As with $\nu$MG-LPA, $\nu$BM-LPA we apply a manual gradient descent to optimize the parameters. However, we arrive at the same parameter values as $\nu$MG-LPA, i.e., a degree threshold $D_H$ of $128$, and kernel launch configurations of $32$ threads per block for the thread-per-vertex kernel and $256$ threads per block for the block-per-vertex kernel. Additionally, $\nu$BM-LPA mitigates community swaps in the same way as $\nu$MG-LPA.

Like $\nu$MG-LPA, $\nu$BM-LPA also has a space complexity of $O(|V|)$ and time complexity of $O(K|E|)$. The psuedocode of $\nu$BM-LPA is given in Algorithm \ref{alg:rakbm}, with its detailed explanation in Section \ref{sec:explain-rakbm}.

\section{Evaluation}
\label{sec:evaluation}
\subsection{Experimental Setup}
\label{sec:setup}

\subsubsection{System used}
\label{sec:system}

We use a server with a 64-core AMD EPYC-7742 processor running at $2.25$ GHz, and NVIDIA A100 GPU which has $80$ GB of global memory ($1935$ GB/s bandwidth), $164$ KB of shared memory/SM, 108 SMs, and 64 CUDA cores/SM. The server also has $512$ GB of DDR4 RAM, and runs Ubuntu 20.04. For CPU-only LPA evaluations, we use a separate server with two 16 core Intel Xeon Gold 6226R processors running at $2.90$ GHz. Each core has $1$ MB L1 cache, $16$ MB L2 cache, and a $22$ MB shared L3 cache. This system also has $512$ GB of RAM and runs CentOS Stream 8.

\subsubsection{Configuration}
\label{sec:configuration}

We use 32-bit integers for vertex IDs, community/label IDs, and sketch keys/labels, and use 32-bit floating-point numbers for edge weights and sketch values. In our GPU implementation of LPA with weighted MG sketches, $\nu$MG8-LPA, we use $8$ slots per sketch \cite{sahu2024memory}, avoid rescanning the top-$k$ community labels, utilize warp-level voting functions, and employ a merge-based kernel (where each thread group creates a sketch for a vertex, which is later merged). For both $\nu$MG8-LPA and our weighted BM algorithm implementation, $\nu$BM-LPA, vertices are split into low and high-degree sets: degrees below $128$ are low, and the rest are high. For low-degree vertices, $\nu$BM-LPA uses a thread-per-vertex kernel with a thread block size of $32$, while high-degree vertices use a block-per-vertex kernel with a thread block size of $256$, where threads collectively find the majority community label using shared memory. For $\nu$MG8-LPA, low-degree vertices are processed with a group-per-vertex kernel with a thread block size of $32$, where the $32$ threads are split into 4 cooperative groups of $8$ threads each. High-degree vertices use a block-per-vertex kernel with a thread block size of $256$, where the $256$ threads are divided into $32$ cooperative groups of $8$ threads each --- after scanning, these thread groups merge their local sketches into a single weighted MG sketch after scanning all edges. Both algorithms use a Pick-Less (PL) setting of $8$ to reduce frequent label swaps, only allowing label changes to lower-ID labels every $8$ iterations --- starting from the first iteration. Further, we use an iteration tolerance of $\tau = 0.05$ and cap iterations at \textit{\small{MAX\_ITERATIONS}} $= 20$ \cite{sahu2023gvelpa}. For compilation we use the \texttt{-O3} optimization flag, and employ CUDA 11.4 on the GPU system. On the CPU-only system, we rely on GCC 8.5 and OpenMP 4.5. All multicore implementations of LPA are executed with 64 threads.

\subsubsection{Dataset}
\label{sec:dataset}

The graphs used in our experiments, shown in Table \ref{tab:dataset}, are from the SuiteSparse Matrix Collection \cite{suite19}. These graphs range from $3.07$ million to $214$ million vertices and $25.4$ million to $3.80$ billion edges. All edges are undirected and weighted, with a default weight of $1$. We did not use publicly available real-world weighted graphs due to their smaller size, although our parallel algorithms can handle weighted graphs without changes. We also exclude SNAP datasets with ground-truth communities, as they are non-disjoint, while our focus is on disjoint communities. It is worth noting that community detection is not just about matching ground truth, which may not accurately reflect a network's real structure and could miss meaningful patterns \cite{peel2017ground}.

\begin{table}[hbtp]
  \centering
  \caption{List of $13$ graphs obtained SuiteSparse Matrix Collection \cite{suite19} (directed graphs are marked with $*$). Here, $|V|$ is the number of vertices, $|E|$ is the number of edges (after adding reverse edges), and $D_{avg}$ is the average degree, and $|\Gamma|$ is the number of communities obtained with \textit{$\nu$MG8-LPA}.}
  \label{tab:dataset}
  \begin{tabular}{|c||c|c|c|c|}
    \toprule
    \textbf{Graph} &
    \textbf{\textbf{$|V|$}} &
    \textbf{\textbf{$|E|$}} &
    \textbf{\textbf{$D_{avg}$}} &
    \textbf{\textbf{$|\Gamma|$}} \\
    \midrule
    \multicolumn{5}{|c|}{\textbf{Web Graphs (LAW)}} \\ \hline
    indochina-2004$^*$ & 7.41M & 341M & 41.0 & 385K \\ \hline
    uk-2002$^*$ & 18.5M & 567M & 16.1 & 863K \\ \hline
    arabic-2005$^*$ & 22.7M & 1.21B & 28.2 & 476K \\ \hline
    uk-2005$^*$ & 39.5M & 1.73B & 23.7 & 1.55M \\ \hline
    webbase-2001$^*$ & 118M & 1.89B & 8.6 & 12.7M \\ \hline
    it-2004$^*$ & 41.3M & 2.19B & 27.9 & 1.50M \\ \hline 
    sk-2005$^*$ & 50.6M & 3.80B & 38.5 & 633K \\ \hline
    \multicolumn{5}{|c|}{\textbf{Social Networks (SNAP)}} \\ \hline
    com-LiveJournal & 4.00M & 69.4M & 17.4 & 175K \\ \hline
    com-Orkut & 3.07M & 234M & 76.2 & 1.91K \\ \hline
    \multicolumn{5}{|c|}{\textbf{Road Networks (DIMACS10)}} \\ \hline
    asia\_osm & 12.0M & 25.4M & 2.1 & 2.86M \\ \hline
    europe\_osm & 50.9M & 108M & 2.1 & 8.04M \\ \hline
    \multicolumn{5}{|c|}{\textbf{Protein k-mer Graphs (GenBank)}} \\ \hline
    kmer\_A2a & 171M & 361M & 2.1 & 41.5M \\ \hline
    kmer\_V1r & 214M & 465M & 2.2 & 50.4M \\ \hline
  \bottomrule
  \end{tabular}
\end{table}

\begin{figure*}[hbtp]
  \centering
  \subfigure[Runtime in seconds (logarithmic scale) with \textit{NetworKit LPA}, \textit{GVE-LPA}, \textit{$\nu$-LPA}, \textit{$\nu$MG8-LPA}, and \textit{$\nu$BM-LPA}]{
    \label{fig:compare--runtime}
    \includegraphics[width=0.98\linewidth]{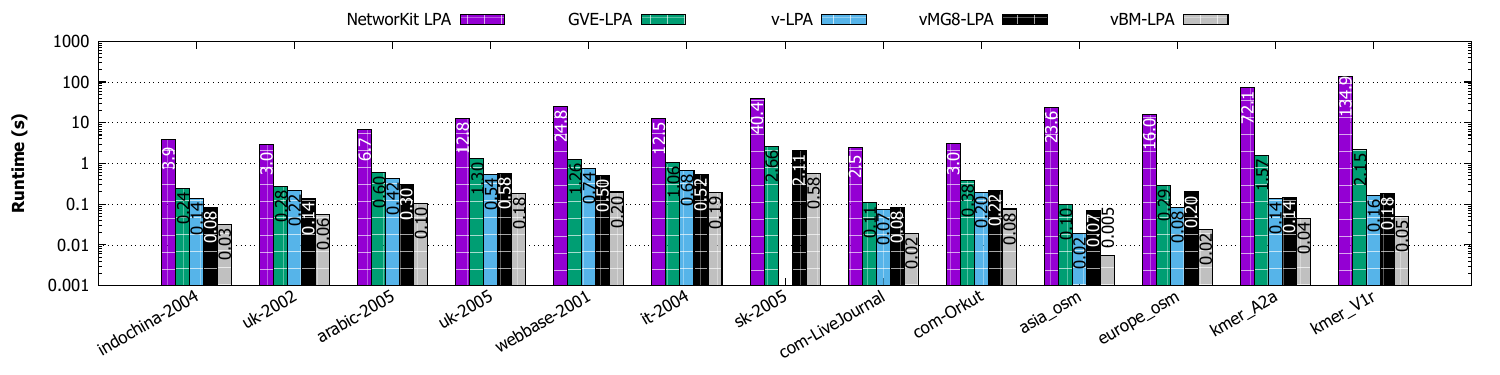}
  }
  \subfigure[Speedup of \textit{$\nu$MG8-LPA} (logarithmic scale) with respect to \textit{NetworKit LPA}, \textit{GVE-LPA}, \textit{$\nu$-LPA}, and \textit{$\nu$BM-LPA}.]{
    \label{fig:compare--speedup}
    \includegraphics[width=0.98\linewidth]{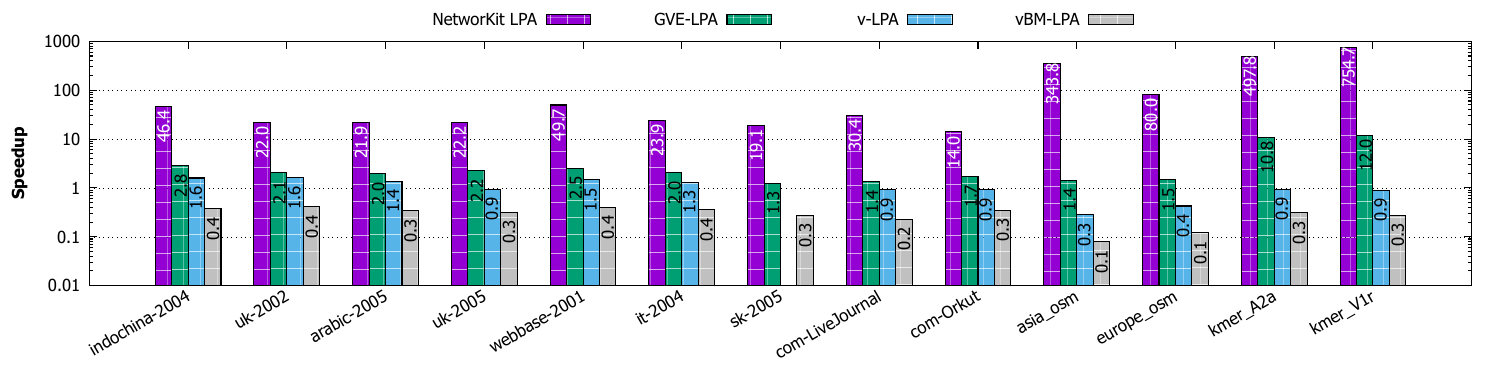}
  }
  \subfigure[Modularity of communities obtained with \textit{NetworKit LPA}, \textit{GVE-LPA}, \textit{$\nu$-LPA}, \textit{$\nu$MG8-LPA}, and \textit{$\nu$BM-LPA}.]{
    \label{fig:compare--modularity}
    \includegraphics[width=0.98\linewidth]{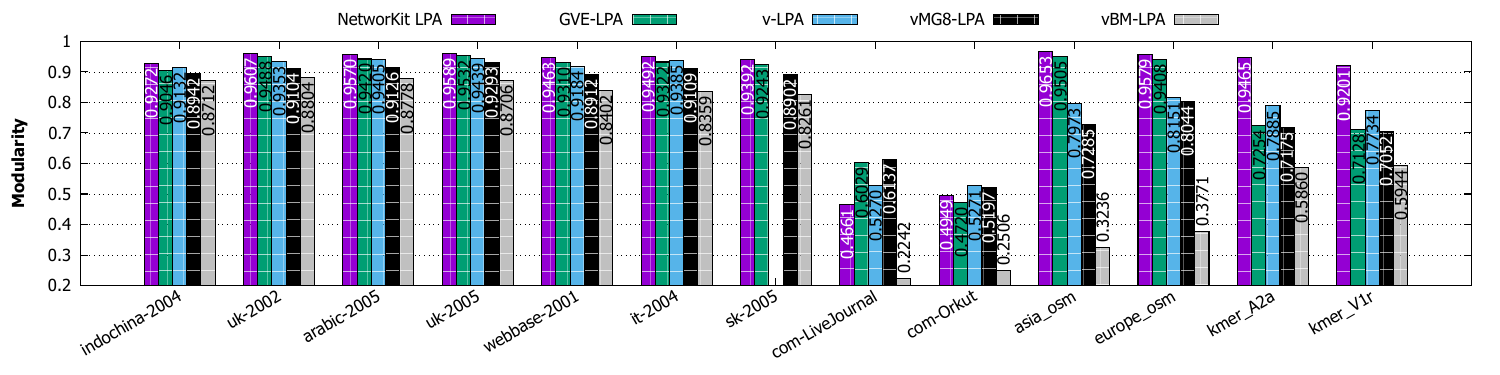}
  }
  \subfigure[Memory usage in gigabytes of \textit{NetworKit LPA}, \textit{GVE-LPA}, \textit{$\nu$-LPA}, \textit{$\nu$MG8-LPA}, and \textit{$\nu$BM-LPA}.]{
    \label{fig:compare--memory}
    \includegraphics[width=0.98\linewidth]{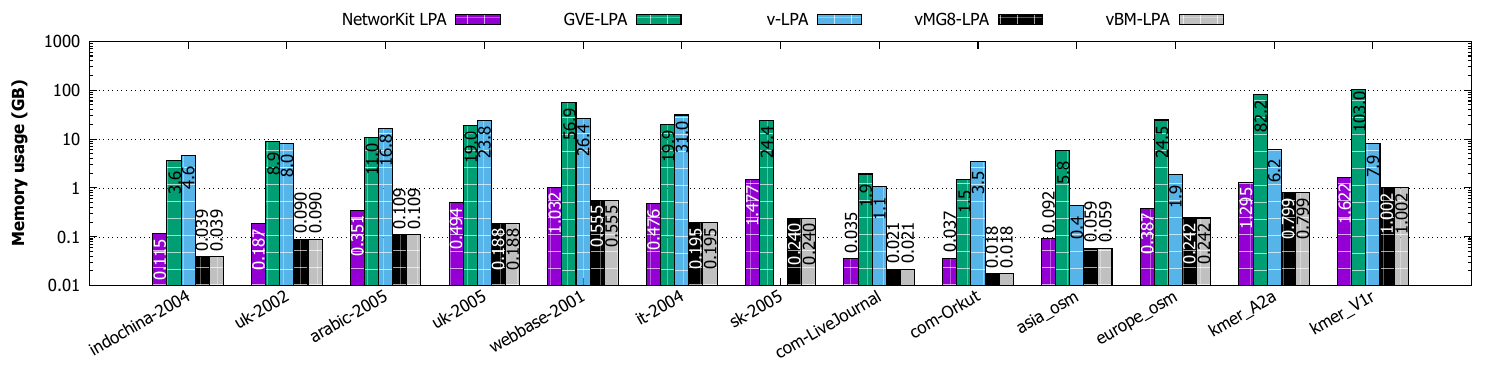}
  } \\[-2ex]
  \caption{Runtime in seconds (log-scale), speedup (log-scale), modularity of obtained communities, and memory usage in gigabytes (log-scale) with \textit{NetworKit LPA}, \textit{GVE-LPA}, \textit{$\nu$-LPA}, \textit{$\nu$MG8-LPA}, and \textit{$\nu$BM-LPA} for each graph in the dataset.}
  \label{fig:compare}
\end{figure*}

\subsection{Performance Comparison}
\label{sec:performance-comparison}  

We evaluate the performance of our algorithms, $\nu$MG8-LPA and $\nu$BM-LPA, in comparison with NetworKit LPA \cite{staudt2016networkit}, GVE-LPA \cite{sahu2023gvelpa}, and $\nu$-LPA \cite{sahu2024nulpa}. NetworKit LPA and GVE-LPA are parallel multicore implementations, while $\nu$-LPA is GPU-based. For NetworKit LPA, we use a Python script to run \texttt{PLP} (Parallel Label Propagation) and measure total runtime with \texttt{getTiming()}. To measure memory usage, we monitor the Resident Set Size (RSS) before running \texttt{PLP} and the peak memory usage during execution by repeatedly reading \texttt{/proc/self/status}. Note that NetworKit LPA might use more memory than reported, as it allocates several small buffers which are likely to have been already reserved by the runtime (from the OS). In contrast, our implementations use large, contiguous memory blocks. For GVE-LPA and $\nu$-LPA, we run their respective scripts. We measure GVE-LPA’s memory usage by checking the RSS before and after memory allocation. For $\nu$-LPA, we use \texttt{cudaMemGetInfo()} to measure memory before and after allocation. We exclude memory used to store the input graph, focusing only on memory used by the algorithm itself, including community labels. Neither GVE-LPA nor $\nu$-LPA allocate additional memory during iterations, so repeated memory tracking during algorithm execution is unnecessary. We perform five runs per graph to calculate average runtimes and modularity of detected communities for each implementation.

Figure \ref{fig:compare--runtime} compares the runtimes of NetworKit LPA, GVE-LPA, $\nu$-LPA, $\nu$MG8-LPA, and $\nu$BM-LPA across different graphs. Figure \ref{fig:compare--speedup} highlights the speedup of $\nu$MG8-LPA relative to other methods. Figure \ref{fig:compare--modularity} displays the modularity scores of the detected communities, while Figure \ref{fig:compare--memory} shows the memory usage of each method (excluding storage for the input graph). Due to an out-of-memory error, $\nu$-LPA results for the \textit{sk-2005} graph are omitted. In terms of memory usage, both $\nu$MG8-LPA and $\nu$BM-LPA achieve, on average, $2.2\times$, $98\times$, and $44\times$ lower memory usage than NetworKit LPA, GVE-LPA, and $\nu$-LPA. Note how this allows $\nu$MG8-LPA and $\nu$BM-LPA to successfully run on the \textit{sk-2005} graph. Further, on average, $\nu$BM-LPA is $186\times$, $9.0\times$, $3.5\times$, and $3.7\times$ faster than NetworKit LPA, GVE-LPA, $\nu$-LPA, and $\nu$MG8-LPA, respectively, but its community quality is $27\%$, $24\%$, $23\%$, and $20\%$ lower than those methods, respectively. In comparison, $\nu$MG8-LPA is $51\times$ and $2.4\times$ faster than NetworKit LPA and GVE-LPA, but $1.1\times$ and $3.7\times$ slower than $\nu$-LPA and $\nu$BM-LPA. It identifies communities that are $8.4\%$, $4.7\%$, and $2.9\%$ lower in quality than NetworKit LPA, GVE-LPA, and $\nu$-LPA, but $25\%$ higher than $\nu$BM-LPA. In particular, we observe that $\nu$MG8-LPA identifies communities of high-quality on web graphs and social networks, but yields lower-quality communities on road networks and protein k-mer graphs. In contrast, $\nu$BM-LPA obtains communities of moderate-quality on web graphs but performs poorly on the other graph types. We plan to address this discrepancy in future work. Despite this, the current findings indicate that $\nu$MG8-LPA is a strong candidate for web graphs and social networks. For road networks, however, GVE-LPA proves to be the most effective, while NetworKit LPA is recommended for protein k-mer graphs. If performance is crucial, $\nu$BM-LPA may be considered for web graphs.


\section{Conclusion}
\label{sec:conclusion}
In summary, this report presents a memory-efficient, GPU-based implementation of the Label Propagation Algorithm (LPA) for community detection, addressing the high memory demands of previous methods like GVE-LPA and $\nu$-LPA. By using weighted Boyer-Moore (BM) and Misra-Gries (MG) sketches, we reduce memory usage without sacrificing performance. The proposed algorithms, $\nu$MG8-LPA and $\nu$BM-LPA, use $2.2\times$, $98\times$, and $44\times$ less memory than NetworKit LPA, GVE-LPA, and $\nu$-LPA, respectively. Further, $\nu$BM-LPA is $186\times$, $9.0\times$, $3.5\times$, and $3.7\times$ faster than these methods but results in lower community quality (up to $27\%$ less). $\nu$MG8-LPA is $51\times$ and $2.4\times$ faster than NetworKit LPA and GVE-LPA, with only a small quality decrease (up to $8.4\%$) compared to these methods. It performs best on web graphs and social networks, while $\nu$BM-LPA is faster on web graphs but less effective on other graph types.

The reduced working set of our algorithms align with the principles of external memory algorithms, where reduced\ignore{in-memory} data transfer enables the handling of larger datasets. When leveraged with unified memory \cite{harris2017unified} to store the input graph, we hope our algorithms facilitate efficient processing of massive graphs on shared memory systems. The next stages of research could focus on improving the community quality tradeoffs, optimizing CUDA configurations for enhanced performance, and exploring FPGA implementations.

\begin{acks}
I would like to thank Prof. Kishore Kothapalli, Prof. Dip Sankar Banerjee, and Balavarun Pedapudi for their support.
\end{acks}

\bibliographystyle{ACM-Reference-Format}
\bibliography{main}

\clearpage
\appendix
\section{Appendix}
\label{sec:appendix}
\subsection{Our Weighted Misra-Gries (MG) based GPU Implementation of LPA}
\label{sec:explain-rakmg}

Algorithm \ref{alg:rakmg} outlines the pseudocode for our GPU-based implementation of LPA using the weighted Misra-Gries (MG) heavy hitters algorithm, which we call $\nu$MG-LPA. Here, the \texttt{lpa()} function takes a graph $G$ as input and outputs the labels $C$ for each vertex.

In \texttt{lpa()}, we start by assigning each vertex a unique community label, setting $C[i]$ to the vertex ID (line \ref{alg:rakmg--init}). We then run LPA iterations up to a maximum of \textit{\small{MAX\_ITERATIONS}} (line \ref{alg:rakmg--iters-begin}). Every $\rho$ iterations, we activate PL mode (line \ref{alg:rakmg--pickless}) to reduce ineffective label swaps. Next, in each iteration, we invoke \texttt{lpaMove()}, which updates the community labels based on local neighborhood information (line \ref{alg:rakmg--move}) and returns $\Delta N$, the number of vertices with changed labels. If the fraction of changed vertices $\Delta N / N$ falls below the specified tolerance $\tau$ and PL mode is inactive, the algorithm has converged, and thus terminates (line \ref{alg:rakmg--converged}). Otherwise, the process repeats until convergence. Finally, the community labels $C$ are returned (line \ref{alg:rakmg--return}).

\begin{algorithm}[hbtp]
\caption{$\nu$MG-LPA: Our GPU-based implementation of LPA, based on weighted Misra-Gries (MG) heavy hitters algorithm.}
\label{alg:rakmg}
\begin{algorithmic}[1]
\Require{$G(V, E)$: Input graph}
\Require{$C$: Community label of each vertex}
\Ensure{$N$: Number of vertices in $G$, i.e., $|V|$}
\Ensure{$S(S_k, S_v)$: Labels, weights array of the MG sketch}
\Ensure{$c^@$: Most weighted candidate label in the sketch}
\Ensure{$\Delta N$: Number of changed vertices, overall}
\Ensure{$\Delta N_G$: Changed vertices per thread group/block}
\Ensure{$R_H$: Number of thread groups processing a vertex}
\Ensure{$k$: Number of slots in the MG sketch}
\Ensure{$s$: Slot index for the current thread}
\Ensure{$g$: Current thread group ID}
\Ensure{$t$: Current thread ID}
\Ensure{$\rho$: Iteration gap for pick-less mode}
\Ensure{$\tau$: Iteration tolerance}

\Statex

\Function{lpa}{$G$}
  \State $C \gets [0 .. |V|)$ \label{alg:rakmg--init}
  \ForAll{$l_i \in [0\ \dots\ \text{\small{MAX\_ITERATIONS}})$} \label{alg:rakmg--iters-begin}
    \State $\rhd$ Mitigate community swaps with \textbf{pick-less}
    \If{$l_i \bmod \rho = 0$} employ \textbf{pick-less} mode\label{alg:rakmg--pickless}
    \EndIf
    \State $\Delta N \gets lpaMove(G, C)$ \label{alg:rakmg--move}
    \If{\textbf{not pick-less and} $\Delta N / N < \tau$} \textbf{break}\label{alg:rakmg--converged}
    \EndIf
  \EndFor \label{alg:rakmg--iters-end}
  \Return{$C$} \label{alg:rakmg--return}
\EndFunction

\Statex

\Function{lpaMove}{$G, C$} \label{alg:rakmgmove--begin}
  \State $S_k \gets \{\{\}\}$ \textbf{;} $S_v \gets \{\{\}\}$ \label{alg:rakmgmove--init-begin}
  \State $\Delta N \gets 0$ \textbf{;} $\Delta N_G \gets \{0\}$
  \State $s \gets t \bmod k$ \textbf{;} $g \gets \lfloor t / k \rfloor$ \textbf{on each thread}\label{alg:rakmgmove--init-end}
  \ForAll{\textbf{unprocessed} $i \in V$ \textbf{in parallel}} \label{alg:rakmgmove--iter-begin}
    \State $\rhd$ Scan communities connected to vertex $i$
    \State $sketchClear(S[g])$ \textbf{in parallel}
    \ForAll{$(j, w) \in G.neighbors(i)$ \textbf{in parallel}} \label{alg:rakmgmove--scan-begin}
      \If{$j = i$} \textbf{continue}
      \EndIf
      \State $sketchAccumulate(S[g], C[j], w, s)$ \textbf{in parallel}
    \EndFor \label{alg:rakmgmove--scan-end}
    \State $\rhd$ Merge multiple sketches into one
    \If{$R_H > 2$} use \textbf{shared} mode below \label{alg:rakmgmove--merge-begin}
    \EndIf
    \If{$g > 0$} \textbf{in parallel}
      \ForAll{$s \in [0\ \dots\ k)$}
        \State $c \gets S_k[g, s]$ \textbf{;} $w \gets S_v[g, s]$
        \If{$w = 0$} \textbf{continue}
        \EndIf
        \State $sketchAccumulate(S[0], c, w, s)$ \textbf{in parallel}
      \EndFor
    \EndIf \label{alg:rakmgmove--merge-end}
    \State $\rhd$ Find best community label for vertex $i$
    \State $c^@ \gets sketchMaxKey(S[0])$ \textbf{in parallel} \label{alg:rakmgmove--maxkey}
    \State $\rhd$ Change label of vertex $i$ to most weighted label $c^@$
    \If{$c^@ \neq C[i]$ \textbf{and} $($\textbf{not pick-less or} $c^@ < C[i])$} \label{alg:rakmgmove--move-begin}
      \State $C[i] \gets c^@$ \textbf{;} $\Delta N_G[g] \gets \Delta N_G[g] + 1$
      \ForAll{$j \in G.neighbors(i)$ \textbf{in parallel}}
        \State Mark $j$ as unprocessed
      \EndFor
    \EndIf \label{alg:rakmgmove--move-end}
  \EndFor \label{alg:rakmgmove--iter-end}
  \State $atomicAdd(\Delta N, \Delta N_G[g])$ \textbf{in parallel} \label{alg:rakmgmove--changed}
  \Return{$\Delta N$} \label{alg:rakmgmove--return}
\EndFunction \label{alg:rakmgmove--end}
\end{algorithmic}
\end{algorithm}

Each iteration of the LPA is handled by the \texttt{lpaMove()} function (line \ref{alg:rakmgmove--begin}). In this function, the community label of each unprocessed vertex $i$ in the graph $G$ is updated. To do this, each vertex $i$ is assigned one or more thread groups based on its degree. A thread group contains exactly $k$ threads, with each thread being responsible for a specific slot in the MG sketch. Each thread's slot index  --- the slot it is responsible for --- is calculated as $s = t \bmod k$, where $t$ is the thread ID, and $k$ is both the number of slots in the sketch and the number of threads in the group. Additionally, each thread group has a unique ID $g = \lfloor t/k \rfloor$. At the start of \texttt{lpaMove()}, we initialize the MG sketch arrays for labels $S_k$ and weights $S_v$, in addition to the overall count $\Delta N$ of changed vertices, and the counts $\Delta N_G$ of changed vertices for each thread group (lines \ref{alg:rakmgmove--init-begin}-\ref{alg:rakmgmove--init-end}). Each vertex $i$ in $G$ is then processed in parallel (line \ref{alg:rakmgmove--iter-begin}), starting with a scan of its neighboring communities to determine the top-$k$ weighted labels. For this, each thread group $g$ clears its private sketch $S[g]$ and then accumulates labels from the vertex's neighbors $j \in J_i$ based on edge weights $w = w_{ij}$ using the \texttt{sketchAccumulate()} function (lines \ref{alg:rakmgmove--scan-begin}-\ref{alg:rakmgmove--scan-end}). The psuedocode of \texttt{sketchAccumulate()} is given in Algorithm \ref{alg:sketch}. After the neighborhood scan, $N_V/k$ sketches, one from each thread group, have now been populated. Here, $N_V$ is the total number of threads per vertex, and $N_V/k$ is the number of thread groups assigned to that vertex. We now proceed to merge these sketches into a single, consolidated sketch in $S[0]$. For this, all thread groups except the first thread group assigned to each vertex ($g > 0$), start to accumulate their top-$k$ identified labels into the sketch $S[0]$ belonging to the first thread group ($g = 0$) of each vertex in parallel (lines \ref{alg:rakmgmove--merge-begin}-\ref{alg:rakmgmove--merge-end}). If more than two thread groups handle a vertex ($N_V/k > 2$), merging is done in ``shared'' mode, using appropriate atomic operations to manage shared updates. An alternative approach is to use a single shared sketch for each vertex $i$, accessible by all thread groups. This requires atomic operations due to concurrent access, which can lead to increased contention. Algorithm \ref{alg:rakmgnm} shows the psuedocode of this approach. Despite avoiding the overhead of merging multiple sketches, this shared approach has shown lower performance, as shown in Section \ref{sec:mg-sketch-merge}. Therefore, we employ the multi-sketch merging approach.

\begin{algorithm}[hbtp]
\caption{Accumulating a label, and its associated weight, in a weighted Misra-Gries (MG) sketch --- using warp-level primitives.}
\label{alg:sketch}
\begin{algorithmic}[1]
\Require{$S(S_k, S_v)$: Labels, weights array of the MG sketch}
\Require{$c, w$: Label, weight to accumulate into the MG sketch}
\Require{$s$: Slot index for the current thread}
\Ensure{$has$: MG sketch has label $c$?}
\Ensure{$s_\phi$: Free slot index}

\Statex

\Function{sketchAccumulate}{$S, c, w, s$}
  \State $\rhd$ Add edge weight to community label
  \If{$S_k[s] = c$} \label{alg:sketch--accumulatewt-begin}
    \If{\textbf{not shared}} $S_v[s] \gets S_v[s] + w$
    \Else\ $atomicAdd(S_v[s], w)$
    \EndIf
  \EndIf \label{alg:sketch--accumulatewt-end}
  \State $has \gets groupBallot(S_k[s], c)$ \label{alg:sketch--accumulated-ballot}
  \State $\rhd$ Done if label is already in the sketch
  \If{$has \neq 0$} \ReturnInline{$done$} \label{alg:sketch--accumulated-done}
  \EndIf
  \State $\rhd$ Find and empty slot, and populate it
  \State $\rhd$ Retry if some other thread reserved the free slot
  \Repeat \label{alg:sketch--populateempty-begin}
    \State $\rhd$ Find empty slot
    \State $B_\phi \gets groupBallot(S_v[s] = 0)$ \label{alg:sketch--findempty-begin}
    \State $s_\phi \gets findFirstSetBit(B_\phi) - 1$
    \If{$B_\phi = 0$} \textbf{break}
    \EndIf \label{alg:sketch--findempty-end}
    \State $\rhd$ Add community label to sketch
    \If{$s_\phi = s$} \label{alg:sketch--populate-begin}
      \If{\textbf{not shared}}
        \State $S_k[s] \gets c$
        \State $S_v[s] \gets w$
      \Else
        \If{$atomicCAS(S_v[s], 0, w) = 0$} $S_k[s] \gets c$
        \Else\ $B_\phi \gets 0$
        \EndIf
      \EndIf
    \EndIf \label{alg:sketch--populate-end}
    \State $\rhd$ $B_\phi$ may have been updated
    \If{\textbf{is shared}} $B_\phi \gets groupAll(B_\phi \neq 0)$ \label{alg:sketch--populatecheck}
    \EndIf
  \Until{\textbf{not shared or} $B_\phi \neq 0$} \label{alg:sketch--populateempty-end}
  \State $\rhd$ Subtract edge weight from non-matching labels
  \If{$B_\phi = 0$} \label{alg:sketch--subtract-begin}
    \If{\textbf{not shared}} $S_v[s] \gets S_v[s] - w$
    \Else\ $atomicAdd(S_v[s], -w)$
    \EndIf
  \EndIf \label{alg:sketch--subtract-end}
  \Return{$done$} \label{alg:sketch--return}
\EndFunction




\end{algorithmic}
\end{algorithm}

After merging the MG sketches from each thread group into a single consolidated sketch for vertex $i$, we identify the most weighted candidate label $c^@$ in the sketch (line \ref{alg:rakmgmove--maxkey}). We do not perform a rescan to find the label with the highest weight for $i$ because it does not improve performance or community quality, as discussed in Section \ref{sec:mg-rescan}. Algorithm \ref{alg:rakmgnm} shows the pseudocode for \texttt{lpaMove()}, where rescan can be used to check the total weight of top labels among $i$'s neighbors. Next, we check if $c^@$ differs from $i$'s current label and if it meets the conditions set by the PL mode (e.g., $c^@$ is smaller than $C[i]$ if PL mode is active). If it does, we update $i$'s label to $c^@$, adjust the count of changed vertices for the current thread group $\Delta N_G$ (noting that only the first thread group updates this count when multiple groups manage the same vertex), and mark all neighboring vertices of $i$ as unprocessed to allow label updates to propagate (lines \ref{alg:rakmgmove--move-begin}-\ref{alg:rakmgmove--move-end}). After all vertices are processed, the count of changed vertices from each thread group $\Delta N_G$ is summed into a global count $\Delta N$ using atomic addition (line \ref{alg:rakmgmove--changed}). Finally, we return $\Delta N$ (line \ref{alg:rakmgmove--return}), allowing the main loop in \texttt{lpa()} to determine if the algorithm has converged or should continue iterating.

\subsection{Populating Misra-Gries (MG) sketch}
\label{sec:explain-sketch}

Algorithm \ref{alg:sketch} presents the pseudocode for accumulating a label and its associated weight into a weighted Misra-Gries (MG) sketch, using warp-level primitives \cite{lin2018using}. Here, the \texttt{sketchAccumulate()} function takes as input a sketch $S$ represented by label $S_k$ and weight $S_v$ arrays, a label $c$, a weight $w$ to be accumulated, and a slot index $s$ in the sketch, that is specific to the current thread.

In the algorithm, we start by checking whether the current slot $s$ already holds the target label $c$. If it does, the weight $w$ is added to the current weight stored in $S_v[s]$ (lines \ref{alg:sketch--accumulatewt-begin}--\ref{alg:sketch--accumulatewt-end}). In ``shared'' mode, where the sketch is shared among multiple thread groups, this addition is performed atomically. If the label $c$ is already present, the \texttt{groupBallot()} function is used to broadcast $c$'s presence across threads within the warp, updating the bits in $has$ accordingly. If $has \neq 0$, indicating the label is found, no further action is needed since the sketch already contains the label (line \ref{alg:sketch--accumulated-done}). If the label is not found, we proceed to locate an empty slot in the sketch to store $c$ and $w$. This search is performed iteratively until a free slot is successfully reserved. Here, we use the \texttt{groupBallot()} function to identify free slots by checking for zero-valued weights (lines \ref{alg:sketch--findempty-begin}--\ref{alg:sketch--findempty-end}), and use the \texttt{findFirstSetBit()} function to determine the first available slot. Once a slot is identified, we attempt to populate it (lines \ref{alg:sketch--populate-begin}--\ref{alg:sketch--populate-end}). In \textit{non-shared} mode, where the sketch is exclusive to a single thread group, the slot is directly assigned. In \textit{shared} mode, an atomic compare-and-swap operation ensures that the assignment only occurs if the slot is still free, preventing race conditions. If another thread claims the slot simultaneously, the search process is restarted. If no empty slot is found ($B_\phi = 0$), weight adjustment is performed in order to maintain the MG sketch. In this case, $w$ is subtracted uniformly from the weights of all existing labels (lines \ref{alg:sketch--subtract-begin}--\ref{alg:sketch--subtract-end}). Finally, the algorithm returns a $done$ status (line \ref{alg:sketch--return}).

\subsection{Our Weighted Boyer-Moore (BM) based GPU Implementation of LPA}
\label{sec:explain-rakbm}

Algorithm \ref{alg:rakbm} presents the pseudocode for our GPU implementation of LPA, which we refer to as $\nu$BM-LPA. This method leverages the weighted Boyer-Moore (BM) majority voting algorithm. The main function of the algorithm is \texttt{lpa()}. It takes a graph $G$ as input and outputs the set of community labels $C$ assigned to each vertex.

\begin{algorithm}[hbtp]
\caption{$\nu$BM-LPA: Our GPU-based implementation of LPA, based on weighted Boyer-Moore (BM) majority vote algorithm.}
\label{alg:rakbm}
\begin{algorithmic}[1]
\Require{$G(V, E)$: Input graph}
\Require{$C$: Community label of each vertex}
\Ensure{$N$: Number of vertices in $G$, i.e., $|V|$}
\Ensure{$c^\#$: Majority weighted label for vertex $i$}
\Ensure{$\Delta N$: Number of changed vertices, overall}
\Ensure{$\Delta N_G$: Changed vertices per thread group}
\Ensure{$g$: Current thread group ID}
\Ensure{$\rho$: Iteration gap for pick-less mode}
\Ensure{$\tau$: Iteration tolerance}

\Statex

\Function{lpa}{$G$}
  \State $C \gets [0 .. |V|)$ \label{alg:rakbm--init}
  \ForAll{$l_i \in [0\ \dots\ \text{\small{MAX\_ITERATIONS}})$} \label{alg:rakbm--iters-begin}
    \State $\rhd$ Mitigate community swaps with \textbf{pick-less}
    \If{$l_i \bmod \rho = 0$} employ \textbf{pick-less} mode\label{alg:rakbm--pickless}
    \EndIf
    \State $\Delta N \gets lpaMove(G, C)$ \label{alg:rakbm--move}
    \If{\textbf{not pick-less and} $\Delta N / N < \tau$} \textbf{break}\label{alg:rakbm--converged}
    \EndIf
  \EndFor \label{alg:rakbm--iters-end}
  \Return{$C$} \label{alg:rakbm--return}
\EndFunction

\Statex

\Function{lpaMove}{$G, C$} \label{alg:rakbmmove--begin}
  \State $\Delta N \gets 0$ \textbf{;} $\Delta N \gets \{0\}$ \label{alg:rakbmmove--init}
  \ForAll{\textbf{unprocessed} $i \in V$ \textbf{in parallel}} \label{alg:rakbmmove--iter-begin}
    \State $\rhd$ Find best community label for vertex $i$
    \State $c^\# \gets C[i]$ \textbf{;} $w^\# \gets 0$
    \ForAll{$(j, w) \in G.neighbors(i)$ \textbf{in parallel}} \label{alg:rakbmmove--scan-begin}
      \If{$i = j$} \textbf{continue}
      \EndIf
      \If{$C[j] = c^\#$} $w^\# \gets w^\# + w$
      \ElsIf{$w^\# > w$} $w^\# \gets w^\# - w$
      \Else\ $c^\# \gets C[j]$ \textbf{;} $w^\# \gets w$
      \EndIf
    \EndFor \label{alg:rakbmmove--scan-end}
    \State $\rhd$ Change label of vertex $i$ to majority label $c^\#$
    \If{$c^\# \neq C[i]$ \textbf{and} $($\textbf{not pick-less or} $c^\# < C[i])$} \label{alg:rakbmmove--move-begin}
      \State $C[i] \gets c^\#$ \textbf{;} $\Delta N_G[g] \gets \Delta N_G[g] + 1$
      \ForAll{$j \in G.neighbors(i)$ \textbf{in parallel}}
        \State Mark $j$ as unprocessed
      \EndFor
    \EndIf \label{alg:rakbmmove--move-end}
  \EndFor \label{alg:rakbmmove--iter-end}
  \State $\rhd$ Update number of changed vertices
  \State $atomicAdd(\Delta N, \Delta N_G[g])$ \textbf{in parallel} \label{alg:rakbmmove--changed}
  \Return{$\Delta N$} \label{alg:rakbmmove--return}
\EndFunction \label{alg:rakbmmove--end}
\end{algorithmic}
\end{algorithm}

In the algorithm, we begin by initializing each vertex $i$ in the graph $G$ with a unique label. In particular, we set $C[i]$ to $i$\ignore{. Specifically, each vertex's initial community label $C$ is set to its own index} (line \ref{alg:rakbm--init}). We then perform LPA iterations, up to a maximum number of \textit{\small{MAX\_ITERATIONS}} (line \ref{alg:rakbm--iters-begin}). During these, we periodically enable the Pick-Less (PL) mode (line \ref{alg:rakbm--pickless}) every $\rho$ iterations --- starting from the first iteration --- to reduce the impact of community swaps. Next, in each iteration, we invoke \texttt{lpaMove()} (line \ref{alg:rakbm--move}) to update labels based on the local neighborhood information. If the proportion of changed vertices $\Delta N / N$ falls below the specified tolerance $\tau$ and the PL mode is not active, convergence has been achieved, and we break out of the loop (line \ref{alg:rakbm--converged}). However, if the PL mode is active, it may lead to fewer label updates, which could falsely trigger convergence. Thus, the loop continues during active PL mode. Once convergence is achieved, the final set of community labels is returned (line \ref{alg:rakbm--return}).

Each iteration of LPA is performed in the \texttt{lpaMove()} function (line \ref{alg:rakbmmove--begin}). Here, first, the number of changed vertices $\Delta N$ is initialized to zero, with separate counters for each thread group (line \ref{alg:rakbmmove--init}). Each vertex $i$ is then processed in parallel (line \ref{alg:rakbmmove--iter-begin}) to determine the best community label using a weighted BM majority vote. This involves scanning $i$'s neighbors (lines \ref{alg:rakbmmove--scan-begin}-\ref{alg:rakbmmove--scan-end}), and updating the candidate label $c^\#$ and its weight $w^\#$ to reflect the most frequent neighboring community label by weight. If the majority label $c^\#$ differs from $i$'s current label and satisfies criteria defined by the PL strategy, $i$'s label is updated to $c^\#$ (lines \ref{alg:rakbmmove--move-begin}-\ref{alg:rakbmmove--move-end}). Additionally, all of $i$'s neighbors are marked as unprocessed to ensure label changes propagate in subsequent iterations. Note that each vertex is processed by a thread group, which can be either a single thread or a thread block, depending on the degree of the vertex. When a thread block is used, the threads within it collaborate using shared memory to determine the best label for the vertex. After label updates, the changed counts $\Delta N_G$ from each thread group are combined using atomic addition (line \ref{alg:rakbmmove--changed}). The total count of changed vertices $\Delta N$ is then returned (line \ref{alg:rakbmmove--return}), allowing the main loop in \texttt{lpa()} to decide whether to continue iterating or halt (if convergence has been achieved).

\subsection{Alternative Weighted Misra-Gries (MG) based GPU Implementation of LPA\ignore{(Non-merge based, Supports Rescan)}}
\label{sec:explain-rakmgnm}

Algorithm \ref{alg:rakmgnm} presents a GPU-based Misra-Gries (MG) implementation of LPA that uses a single shared MG sketch per vertex (\textit{non-merge based}) and supports rescanning the top-$k$ weighted labels to determine the most weighted label for each vertex. While this approach does not improve performance, it is included for comparison. Here, as before, the \texttt{lpa()} function takes a graph $G$ as input and outputs the community labels $C$ for each vertex in $G$.

In \texttt{lpa()}, the algorithm starts by assigning each vertex a unique label, setting $C[i]$ to $i$ (line \ref{alg:rakmgnm--init}). It then iterates up to a maximum of \textit{\small{MAX\_ITERATIONS}}, or until convergence (lines \ref{alg:rakmgnm--iters-begin}-\ref{alg:rakmgnm--iters-end}). To mitigate unnecessary label swaps, it switches to Pick-Less (PL) mode every $\rho$ iterations, including the first iteration (line \ref{alg:rakmgnm--pickless}), as earlier. During each iteration, the \texttt{lpaMove()} function updates label assignments (line \ref{alg:rakmgnm--move}). The algorithm stops early if the fraction $\Delta N / N$ of label changes drops below a threshold $\tau$, indicating convergence (line \ref{alg:rakmgnm--converged}).

Each iteration of the LPA is executed in the \texttt{lpaMove()} function (line \ref{alg:rakmgnmmove--begin}), which updates the community label of each unprocessed vertex $i$ in the graph $G$. As earlier, each vertex $i$ is assigned one or more thread groups based on its degree. At the start of \texttt{lpaMove()}, the MG sketch arrays for labels $S_k$ and weights $S_v$ are initialized, along with the total count of changed vertices $\Delta N$ and the counts of changed vertices for each thread group $\Delta N_G$ (lines \ref{alg:rakmgnmmove--init-begin}-\ref{alg:rakmgnmmove--init-end}). Each vertex $i$ in $G$ is then processed in parallel (line \ref{alg:rakmgnmmove--iter-begin}). The process begins with scanning the neighboring communities of vertex $i$ to identify the top-$k$ weighted labels. During this step, the threads first clear the shared sketch $S$. Subsequently, thread groups collaborate on the shared sketch, accumulating labels from the neighbors $j \in J_i$ of vertex $i$ based on edge weights ($w = w_{ij}$) using the \texttt{sketchAccumulate()} function (lines \ref{alg:rakmgnmmove--scan-begin}-\ref{alg:rakmgnmmove--scan-end}). An alternative implementation of \texttt{sketchAccumulate()} that does not use warp-level primitives is provided in Algorithm \ref{alg:sketchnw}. It employs atomic operations to ensure thread-safe updates to the shared sketch $S$. After the neighborhood scan, the shared sketch $S$ is fully populated.

If a rescan is requested (line \ref{alg:rakmgnmmove--rescan-begin}), the algorithm calculates the exact total weight for each of the top-$k$ labels in the sketch $S$ by reexamining $i$'s neighboring vertices, after having cleared the sketch weights (lines \ref{alg:rakmgnmmove--rescan-begin}-\ref{alg:rakmgnmmove--rescan-end}). It then checks if $c^\#$ (the most weighted sub-majority label) differs from $i$'s current label and satisfies the PL mode conditions (e.g., $c^\# < C[i]$ if PL mode is active). If these conditions are met, $i$'s label is updated to $c^\#$, the change count for the thread group $\Delta N_G$ is incremented (only by the first thread group for shared vertices), and $i$'s neighbors are marked unprocessed for further updates (lines \ref{alg:rakmgnmmove--move-begin}-\ref{alg:rakmgnmmove--move-end}). After all vertices are processed, the thread group counts $\Delta N_G$ are combined into a global count $\Delta N$ using atomic addition (line \ref{alg:rakmgnmmove--changed}). The algorithm then returns $\Delta N$\ignore{(line \ref{alg:rakmgnmmove--return}), which determines if the main loop in \texttt{lpa()} should\ignore{continue or} terminate}.

\begin{algorithm}[hbtp]
\caption{A GPU-based implementation of LPA, based on weighted Misra-Gries (MG) heavy hitters algorithm, where all threads update a single shared sketch directly, eliminating the need for a merging step. It also supports rescanning sub-majority labels.}
\label{alg:rakmgnm}
\begin{algorithmic}[1]
\Require{$G(V, E)$: Input graph}
\Require{$C$: Community label of each vertex}
\Ensure{$N$: Number of vertices in $G$, i.e., $|V|$}
\Ensure{$S(S_k, S_v)$: Labels, weights array of the MG sketch}
\Ensure{$c^\#$: Sub-majority weighted label for vertex $i$}
\Ensure{$\Delta N$: Number of changed vertices, overall}
\Ensure{$\Delta N_G$: Changed vertices per thread group/block}
\Ensure{$k$: Number of slots in the MG sketch}
\Ensure{$s$: Slot index for the current thread}
\Ensure{$g$: Current thread group ID}
\Ensure{$t$: Current thread ID}
\Ensure{$\rho$: Iteration gap for pick-less mode}
\Ensure{$\tau$: Iteration tolerance}

\Statex

\Function{lpa}{$G$}
  \State $C \gets [0 .. |V|)$ \label{alg:rakmgnm--init}
  \ForAll{$l_i \in [0\ \dots\ \text{\small{MAX\_ITERATIONS}})$} \label{alg:rakmgnm--iters-begin}
    \State $\rhd$ Mitigate community swaps with \textbf{pick-less}
    \If{$l_i \bmod \rho = 0$} employ \textbf{pick-less} mode\label{alg:rakmgnm--pickless}
    \EndIf
    \State $\Delta N \gets lpaMove(G, C)$ \label{alg:rakmgnm--move}
    \If{\textbf{not pick-less and} $\Delta N / N < \tau$} \textbf{break}\label{alg:rakmgnm--converged}
    \EndIf
  \EndFor \label{alg:rakmgnm--iters-end}
  \Return{$C$} \label{alg:rakmgnm--return}
\EndFunction

\Statex

\Function{lpaMove}{$G, C$} \label{alg:rakmgnmmove--begin}
  \State $S_k \gets \{\}$ \textbf{;} $S_v \gets \{\}$ \label{alg:rakmgnmmove--init-begin}
  \State $\Delta N \gets 0$ \textbf{;} $\Delta N_G \gets \{0\}$
  \State $s \gets t \bmod k$ \textbf{;} $g \gets \lfloor t / k \rfloor$ \textbf{on each thread}\label{alg:rakmgnmmove--init-end}
  \State \textbf{Use shared mode throughout}
  \ForAll{\textbf{unprocessed} $i \in V$ \textbf{in parallel}} \label{alg:rakmgnmmove--iter-begin}
    \State $\rhd$ Scan communities connected to vertex $i$
    \State $sketchClear(S)$ \textbf{in parallel}
    \ForAll{$(j, w) \in G.neighbors(i)$ \textbf{in parallel}} \label{alg:rakmgnmmove--scan-begin}
      \If{$j = i$} \textbf{continue}
      \EndIf
      \State $sketchAccumulate(S, C[j], w, s)$ \textbf{in parallel}
    \EndFor \label{alg:rakmgnmmove--scan-end}
    \State $\rhd$ Rescan sub-majority labels to find the most weighted
    \If{\textbf{rescan requested}} \label{alg:rakmgnmmove--rescan-begin}
      \State $sketchClearValues(S)$ \textbf{in parallel}
      \ForAll{$(j, w) \in G.neighbors(i)$ \textbf{in parallel}}
        \If{$j = i$ \textbf{or} $S_k[s] \neq C[j]$} \textbf{continue}
        \EndIf
        \State $atomicAdd(S_v[s], w)$
      \EndFor
    \EndIf \label{alg:rakmgnmmove--rescan-end}
    \State $\rhd$ Find best community label for vertex $i$
    \State $c^\# \gets sketchMaxKey(S)$ \textbf{in parallel} \label{alg:rakmgnmmove--maxkey}
    \State $\rhd$ Change label of vertex $i$ to most weighted label $c^\#$
    \If{$c^\# \neq C[i]$ \textbf{and} $($\textbf{not pick-less or} $c^\# < C[i])$} \label{alg:rakmgnmmove--move-begin}
      \State $C[i] \gets c^\#$ \textbf{;} $\Delta N_G[g] \gets \Delta N_G[g] + 1$
      \ForAll{$j \in G.neighbors(i)$ \textbf{in parallel}}
        \State Mark $j$ as unprocessed
      \EndFor
    \EndIf \label{alg:rakmgnmmove--move-end}
  \EndFor \label{alg:rakmgnmmove--iter-end}
  \State $atomicAdd(\Delta N, \Delta N_G[g])$ \textbf{in parallel} \label{alg:rakmgnmmove--changed}
  \Return{$\Delta N$} \label{alg:rakmgnmmove--return}
\EndFunction \label{alg:rakmgnmmove--end}
\end{algorithmic}
\end{algorithm}

\begin{algorithm}[hbtp]
\caption{Accumulating a label, and its associated weight, in a weighted Misra-Gries (MG) sketch --- without warp-level primitives.}
\label{alg:sketchnw}
\begin{algorithmic}[1]
\Require{$S(S_k, S_v)$: Labels, weights array of the MG sketch}
\Require{$c, w$: Label, weight to accumulate into the MG sketch}
\Require{$s$: Slot index for the current thread}
\Ensure{$has$: MG sketch has label $c$? / Free slot index}

\Statex

\Function{sketchAccumulate}{$S, c, w, s$}
  \If{$s = 0$} $has \gets -1$
  \EndIf
  \State $\rhd$ Add edge weight to community label
  \If{$S_k[s] = c$} \label{alg:sketchnw--accumulatewt-begin}
    \If{\textbf{not shared}} $S_v[s] \gets S_v[s] + w$
    \Else\ $atomicAdd(S_v[s], w)$
    \EndIf
    \State $has \gets 0$
  \EndIf \label{alg:sketchnw--accumulatewt-end}
  \State $\rhd$ Done if label is already in the list
  \If{$has = 0$} \ReturnInline{$done$} \label{alg:sketchnw--accumulated-done}
  \EndIf
  \State $\rhd$ Find and empty slot, and populate it
  \State $\rhd$ Retry if some other thread reserved the free slot
  \Repeat \label{alg:sketchnw--populateempty-begin}
    \State $\rhd$ Find an empty slot
    \If{$S_v[s] = 0$} $atomicMax(has, s)$ \label{alg:sketchnw--findempty-begin}
    \EndIf
    \If{$has < 0$} \textbf{break}
    \EndIf \label{alg:sketchnw--findempty-end}
    \State $\rhd$ Add community label to list
    \If{$has = s$} \label{alg:sketchnw--populate-begin}
      \If{\textbf{not shared}}
        \State $S_k[s] \gets c$
        \State $S_v[s] \gets w$
      \Else
        \If{$atomicCAS(S_v[s], 0, w) = 0$} $S_k[s] \gets c$
        \Else\ $has \gets 1$
        \EndIf
      \EndIf
    \EndIf \label{alg:sketchnw--populate-end}
  \Until{\textbf{not shared or} $has \geq 0$} \label{alg:sketchnw--populateempty-end}
  \State $\rhd$ Subtract edge weight from non-matching labels
  \If{$has < 0$} \label{alg:sketchnw--subtract-begin}
    \If{\textbf{not shared}} $S_v[s] \gets S_v[s] - w$
    \Else\ $atomicAdd(S_v[s], -w)$
    \EndIf
  \EndIf \label{alg:sketchnw--subtract-end}
  \Return{$done$} \label{alg:sketchnw--return}
\EndFunction
\end{algorithmic}
\end{algorithm}

\subsection{Alternative Method for Populating Misra-Gries (MG) sketch}
\label{sec:explain-sketchnw}

Algorithm \ref{alg:sketchnw} presents a method to update a weighted Misra-Gries (MG) sketch,\ignore{The MG sketch is used to track the frequency of items in a data stream, where each slot can hold a key-value pair.} which does \textit{not use warp-level primitives}. Here, the function \texttt{sketchAccumulate()} takes as input the MG sketch $S$, with labels array $S_k$ and weights array $S_v$, the key-value pair $(c, w)$ to be accumulated, and the current thread's slot index $s$.

At the beginning of the function, if the current thread is responsible for the first slot ($s = 0$), it initializes $has$ to $-1$, indicating that no match has been found yet. The algorithm proceeds to check if the target label $c$ is already present in the sketch. If the slot $s$ holds the same label ($S_k[s] = c$), the corresponding value $S_v[s]$ is updated by adding $w$. This operation is performed atomically if the data is shared among threads. Once the key-value pair is updated, the variable $has$ is set to $0$, indicating that the label was found, and the function returns immediately. If the label is not found, the algorithm attempts to find an empty slot in the sketch where the new key-value pair $(c, w)$ can be inserted. In a loop, each thread checks if its assigned slot is empty ($S_v[s] = 0$). If so, the thread attempts to reserve the slot using an atomic operation, and setting $has$ to the index of the slot if it is successful. If no empty slot is found and $has$ remains negative, the loop exits. Once a free slot is found (i.e., $has$ matches the current thread's slot $s$), the algorithm inserts the key-value pair $(c, w)$. If data is not shared, the assignment is straightforward: $S_k[s] \gets c$ and $S_v[s] \gets w$. If data is shared among threads, the algorithm uses an atomic compare-and-swap operation to safely set the value. If another thread has already reserved the slot, the function retries until the operation succeeds. Finally, if no suitable slot was available for the label (i.e., $has$ remains negative), the algorithm subtracts the weight $w$ from all the slots in the sketch. The subtraction is performed atomically if the sketch is shared. Finally, the function then returns.



\end{document}